\documentclass[sigconf]{acmart}

\AtBeginDocument{%
  \providecommand\BibTeX{{%
    \normalfont B\kern-0.5em{\scshape i\kern-0.25em b}\kern-0.8em\TeX}}}

\usepackage{subfigure} 

\usepackage{bm}
\usepackage{threeparttable}

\setcopyright{acmcopyright}
\copyrightyear{2020} 
\acmYear{2020} 
\acmConference[SIGIR '20]{Proceedings of the 43rd International ACM SIGIR Conference on Research and Development in Information Retrieval}{July 25--30, 2020}{Virtual Event, China}
\acmBooktitle{Proceedings of the 43rd International ACM SIGIR Conference on Research and Development in Information Retrieval (SIGIR '20), July 25--30, 2020, Virtual Event, China}
\acmPrice{15.00}
\acmDOI{10.1145/3397271.3401156}
\acmISBN{978-1-4503-8016-4/20/07}

\begin{document}

%%
%% The "title" command has an optional parameter,
%% allowing the author to define a "short title" to be used in page headers.
\fancyhead{}
%% The "title" command has an optional parameter,
%% allowing the author to define a "short title" to be used in page headers.
%\title[Chunk]{Chunk Accelerates Memory-based Neural Recommendations}
\title{Parameter-Efficient Transfer from Sequential Behaviors for User Modeling and Recommendation
}
%%
%% The "author" command and its associated commands are used to define
%% the authors and their affiliations.
%% Of note is the shared affiliation of the first two authors, and the
%% "authornote" and "authornotemark" commands
%% used to denote shared contribution to the research.
\author{Fajie Yuan}
\affiliation{%
	\institution{Tencent}
	\city{Shenzhen} 
	\state{China} 
}
\email{fajieyuan@tencent.com }

\author{Xiangnan He}
\affiliation{%
	\institution{University of Science and Technology of China}
	\city{Hefei} 
	\state{China} 
}
\email{xiangnanhe@gmail.com}

\author{Alexandros Karatzoglou} 
\affiliation{%
	\institution{Google}
	\city{London} 
	\state{UK} 
}
\email{alexandros.karatzoglou@gmail.com}
\authornote{\scriptsize  Part of this work was done when Alexandros was at Telefonica Research, Spain.}
\author{Liguang Zhang} 
\affiliation{%
	\institution{Tencent}
	\city{Shenzhen} 
	\state{China} 
}
\email{kanongzhang@tencent.com}

%%
%% By default, the full list of authors will be used in the page
%% headers. Often, this list is too long, and will overlap
%% other information printed in the page headers. This command allows
%% the author to define a more concise list
%% of authors' names for this purpose.
%\renewcommand{\shortauthors}{Trovato and Tobin, et al.}

%%
%% The abstract is a short summary of the work to be presented in the
%% article.
	\begin{abstract}
	Inductive transfer learning has had a big impact on computer vision and NLP domains but has not been used in the area of recommender systems. Even though there has been a large body of research on generating recommendations based on modeling user-item interaction sequences, few of them attempt to represent and transfer these models for serving downstream tasks where only limited data exists.
	
	In this paper, we delve on the task of effectively learning a single user representation that can be applied to a diversity of tasks, from  cross-domain recommendations to user profile predictions. 
	Fine-tuning  a large pre-trained network and adapting it to downstream tasks is an effective  way to solve such tasks. However, fine-tuning is parameter inefficient considering that an entire model needs to be re-trained for every new task. To overcome this issue, we  develop a  \underline{p}arameter-\underline{e}fficient  \underline{t}ransf\underline{er} learning architecture, termed as PeterRec, which can be configured on-the-fly to various downstream tasks. Specifically, PeterRec allows the pre-trained parameters to remain unaltered during fine-tuning by injecting a series of re-learned  neural networks, which are small but as expressive as learning
	the entire network.
	We perform extensive experimental ablation to show the effectiveness of the learned user representation in five downstream tasks. Moreover, we  show that 
	PeterRec performs efficient transfer learning in multiple domains, where it achieves  comparable or sometimes better performance relative to fine-tuning the entire model parameters.
	Codes and datasets are available at \color{blue}{\url{https://github.com/fajieyuan/sigir2020\_peterrec}}.
	%	, while only adding  
	%	a fraction of trainable parameters.
	%while only
	% training less than 1\% of them. 
	%	\url{https://github.com/fajieyuan/sigir2020_peterrec}
	%	Codes and datasets are available on: \textit{https://github.com/fajieyuan/sigir2020_peterrec}.
	
\end{abstract}

%%
%% The code below is generated by the tool at http://dl.acm.org/ccs.cfm.
%% Please copy and paste the code instead of the example below.
%%
%\begin{CCSXML}
%<ccs2012>
% <concept>
%  <concept_id>10010520.10010553.10010562</concept_id>
%  <concept_desc>Computer systems organization~Embedded systems</concept_desc>
%  <concept_significance>500</concept_significance>
% </concept>
% <concept>
%  <concept_id>10010520.10010575.10010755</concept_id>
%  <concept_desc>Computer systems organization~Redundancy</concept_desc>
%  <concept_significance>300</concept_significance>
% </concept>
% <concept>
%  <concept_id>10010520.10010553.10010554</concept_id>
%  <concept_desc>Computer systems organization~Robotics</concept_desc>
%  <concept_significance>100</concept_significance>
% </concept>
% <concept>
%  <concept_id>10003033.10003083.10003095</concept_id>
%  <concept_desc>Networks~Network reliability</concept_desc>
%  <concept_significance>100</concept_significance>
% </concept>
%</ccs2012>
%\end{CCSXML}

%%\ccsdesc[500]{Computer systems organization~Embedded systems}
%%\ccsdesc[300]{Computer systems organization~Redundancy}
%%\ccsdesc{Computer systems organization~Robotics}
%%\ccsdesc[100]{Networks~Network reliability}

%%
%% Keywords. The author(s) should pick words that accurately describe
%% the work being presented. Separate the keywords with commas.
	\keywords{Transfer Learning, Recommender System, User Modeling,  Pretraining and Finetuning}
\maketitle

\section{Introduction}
\label{Introduction}
The last 10 years have seen the ever increasing use of social media platforms and e-commerce systems, such as Tiktok, Amazon or Netflix.  Massive amounts of clicking \& purchase interactions, and other user feedback are created explicitly or implicitly in such systems.  For example,  regular users  on Tiktok may watch hundreds to thousands of micro-videos per week given that the average playing time of each video is less than 20 seconds \cite{yuan2020future}. A large body of research has clearly shown that these interaction sequences can be used to model the item preferences of the users  \cite{qu2020cmnrec,yuan2019simple,kang2018self,yuan2020future,yuan2016lambdafm,guo2019dynamic}. Deep neural network models, such as GRURec~\cite{ hidasi2015session} and NextItNet ~\cite{yuan2019simple}, have achieved remarkable results in modeling sequential user-item interactions and generating personalized recommendations.
%Also,  as reported\footnote{\scriptsize \url{https://www.mckinsey.com/industries/retail/our-insights/how-retailers-can-keep-up-with-consumers}}, 35\% of purchase on Amazon and 75\%  of watches on Netflix comes from user behavior analysis and personalized recommendation. 
However, most of the past work has been focused on the task of recommending items on the same platform, from where the data came from. Few of these methods exploit this data to learn a universal user representation that could then be used for a different downstream task, such as for instance the cold-start user problem on a different recommendation platform or the prediction of a user profile.

In this work, we deal with the task of adapting a singe user representation model for multiple downstream tasks. In particular, we attempt to use deep neural network models, pre-trained in an unsupervised (self-supervised) manner on a source domain with rich sequential user-item interactions, for a variety of tasks on target domains, where users are cold or new. To do so, we need to tackle the following issues: (1) construct a highly effective and general pre-training model that is capable of  modeling and representing very long-range user-item interaction sequences without supervision. (2) develop a fine-tuning architecture that can transfer pre-trained user representations to
downstream tasks. Existing recommender systems literature is unclear on  whether unsupervised learned user representations are  useful in different domains where the same users are involved but where they have little supervised labeled data. (3) introduce an adaptation method that enables the fine-tuning architecture to share most of the parameters across all tasks.  
Although fine-tuning a separate model for  each task often performs better, we believe there are important reasons for reusing  parameters between tasks.  Particularly for resource-limited devices, applying  several different neural networks for each task with the same input is computationally expensive and memory intensive~\cite{mudrakarta2018k,stickland2019bert}.  Even for the large-scale web applications,  practitioners need to avoid maintaining a separate large model for every user~\cite{stickland2019bert}, especially when there are a large number of tasks.
%However, to the best of our knowledge, thus far there is no prior work that study parameter sharing problem for transfer or multi-domain learning in the field of recommender systems.  
%(i.e., freezing all  parameters of the pre-trained model)

To tackle the third issue, two transfer techniques have been widely  used ~\cite{yosinski2014transferable}: (1) fine-tuning an additional output layer to project transferred knowledge from a source domain to a target domain,  and (2) fine-tuning the last (few) hidden layers along with the output layer. In fact, we find that fine-tuning only the output layer  often performs poorly in the recommendation scenario;  fine-tuning the last few layers
properly sometimes offers promising performance, but requires much manual effort since the number of layers to be tuned highly depends on the pre-trained model and target task. Thus far, there is no consensus on how to choose the number, which in practice often relies on an inefficient hyper-parameter search. 
%Some research in computer vision~\cite{guo2019spottune,veit2016residual} argues that the last-layer fine-tuning paradigm
%may not theoretically hold for the state-of-the-art multi-path  deep architectures, such as ResNets~\cite{he2016deep}, which have also been widely used in recommendation models~\cite{yuan2019simple,kang2018self,yuan2020future}.
In addition,  fine-tuning the last few layers does not realize our goal to share  most parameters of the pre-trained model. 

%Moreover, revising the pre-trained model for every new task is inflexible for deploying and managing online systems.

To achieve the first two goals, we propose a two-stage training procedure. First, in order to learn a universal user representation, we employ sequential neural networks as our pre-trained model and train them with users' historical clicking or purchase sequences. Sequential models can be trained without manually labeled data using self-supervision which is essentially trained by predicting the next item on the sequence. Moreover sequential data is much easier to collect  from online  systems.
In this paper, we choose NextItNet-style~\cite{yuan2019simple,yuan2020future} neural networks as the base models
considering that they achieve state-of-the-art performance when modeling very long-range sequential user-item interactions~\cite{wang2019towards}. Subsequently, we can adapt the pre-trained model to downstream tasks using  supervised objectives. By doing so, we obtain an NLP~\cite{radford2018improving, houlsby2019parameter} or computer vision (CV)~\cite{yosinski2014transferable,rosenfeld2018incremental}-like transfer learning framework. 
% Figure~\ref{overviewpeter} is an overview of user profile prediction using NextItNet-style sequential neural network.
%Though simple, we find that this work is the first to  evidence that representations unsupervisedly learned from a  sequential recommendation model can largely improve the fine-tuning performance in many downstream tasks,  where only cold or new users are available. 

To achieve the third goal that enables a high degree of parameter sharing for fine-tuning models between domains, we borrow an idea from the learning-to-learn method, analogous to~\cite{bertinetto2016learning}.
The core idea of learning-to-learn is that the parameters of deep neural networks can be predicted from another \cite{rebuffi2017learning,bertinetto2016learning}; moreover, ~\cite{denil2013predicting} demonstrated that
it is possible to predict more than 95\% parameters of a network in a layer given the remaining 5\%. 
%  since learned parameters tend to be structured in computer vision.
Taken inspiration from these works, we are interested in exploring whether these findings hold for the transfer learning tasks in the recommender system (RS) domain. In addition, unlike above works, we are more interested in exploring the idea of parameter adaptation rather than prediction.    Specifically, we propose a separate  grafting neural network, termed as model patch,  which adapts the parameters of each convolutional layer in the pre-trained model to a target task.
%serves as an parameter adapter for each convolutional layer of the pre-trained model. 
Each model patch consists of less than 10\% of the parameters of the original convolutional layer.
% using a residual network architecture.
By inserting such model patches into the pre-trained models, our fine-tuning networks are not only able to keep all  pre-trained parameters unchanged, but also successfully induce them for problems of downstream tasks without a significant drop in performance. We name the proposed model PeterRec, where `Peter' stands for \underline{p}arameter \underline{e}fficient \underline{t}ransf\underline{er} learning.

The contributions of this paper are listed  as follows:
\begin{itemize}
	\item We propose a universal user representational learning architecture, a method that can be used to achieve NLP or CV-like transfer learning for various downstream tasks. More importantly, we are the first to demonstrate  that self-supervised learned user representations can be used to infer user 
	profiles, such as for instance the  gender,  age, preferences and life status (e.g., single, married or parenting). It is conceivable that the inferred user profiles by PeterRec can help improve the quality of many public and commercial services, but also raises concerns of privacy protection.
	\item We propose a simple yet very effective  grafting network, i.e., model patch, which
	 allows pre-trained weights  to remain unaltered and shared for various downstream tasks.
	\item  We propose two alternative ways to inject the model patches into pre-trained models, namely serial and parallel insertion.
	\item  We perform extensive ablation analysis on five different tasks during fine-tuning, and report many insightful findings, which could be directions for future research in the RS domain.
	\item  We have released a high-quality  dataset used for transfer learning research. 
	To our best knowledge, this is the first large-scale  recommendation dataset that can be used for both transfer \& multi-domain learning. 
	We hope our datasets can provide a  benchmark  to facilitate the research of transfer  and multi-domain learning in the RS domain.
\end{itemize}

\section{Related Work}
%Our work intersects with research lines in sequential recommendations, transfer learning and meta learning.
PeterRec tackles two research questions: (1) training an effective and efficient base model, and (2)
transferring the learned user representations from the base model to downstream tasks with a high degree of parameter sharing. Since we choose the sequential  recommendation models to perform this upstream task, we briefly review related literature. Then we recapitalize work in transfer learning and user representation adaptation.
\subsection{Sequential Recommendation Models}

%
%As one of the emerging recommendation problems, sequential recommendations (SR) have been extensively studied in recent years~\cite{hidasi2015session, tang2018caser,yuan2019simple,yuan2020future,guo2019dynamic}. 
A sequential recommendation (SR)  model takes in a sequence (session) of user-item interactions, and taking sequentially each item of the sequence as input aims to predict the next one(s) that the user likes.
SR have demonstrated obvious accuracy  gains compared to traditional content or context-based recommendations when modeling users sequential actions~\cite{kang2018self}. Another merit of SR is that sequential models do not necessarily require user profile information  since user representations can be implicitly reflected  by their past sequential  behaviors. Amongst these models, researchers have paid special attention to  three lines of work: RNN-based \cite{hidasi2015session}, CNN-based~\cite{tang2018caser,yuan2019simple,yuan2020future}, and pure attention-based~\cite{kang2018self} sequential models. In general, typical RNN models strictly rely on  sequential dependencies during training, and thus,  cannot take full advantage of modern computing architectures, such as GPUs or TPU~\cite{yuan2019simple}. 
CNN and attention-based recommendation models do not have such a problems since the entire sequence can be observed during training and thus can be fully parallel. 
%In addition,  CNN-based models also yield better recommendation accuracy since much more neural layers can be  stacked than RNNs. 
%For example, NextItNet-style CNN-based algorithms have shown that the optimal recommendation accuracy is usually accompanied with more than 10 convolutional layers~\cite{yuan2020future}.
%\footnote{\scriptsize In our later work, we find that the performance of NextItNet keeps increasing until it has up to 40 layers on the ML100~\cite{yuan2020future} datasets.} 
One well-known obstacle that prevents CNN from being a strong sequential model is the limited receptive field due to its small kernel size (e.g., 3 $\times$ 3). This  issue has been cleverly approached by introducing  the dilated  convolutional operation, which enables an exponentially increased receptive field with unchanged kernel~\cite{yuan2019simple,yuan2020future}. By contrast, self-attention based sequential models, such as SASRec~\cite{kang2018self} may have  time complexity and memory issues since they grow quadratically with the sequence length.
%Particularly, in our scenario, considering that most users have hundreds or even thousands of news and micro-video watching behaviors in practice, SASRec is not an efficient choice.
%Another downsize is that the attention operation cannot capture the order relationship, and have to seek help from an additional positional embeddings, which is not well-suited to handle variable length sequence prediction problem.
Thereby,  we choose dilated convolution-based sequential neural network to build the pre-trained model by investigating both causal (i.e., NextItNet~\cite{yuan2019simple}) and non-causal (i.e., the bidirectional encoder of GRec~\cite{yuan2020future})  convolutions in this paper.

\subsection{Transfer Learning \& Domain Adaptation}
Transfer learning (TL) has recently become a research hotspot in many application fields of machine learning~\cite{devlin2018bert,rebuffi2018efficient,houlsby2019parameter,radford2018improving}. TL refers to methods that exploit knowledge gained in a source domain where a vast amount of training data is available, to improve a different but related problem in a target domain where only little labeled data can be obtained. Unlike much early work that concentrated on shallow classifiers (or predictors), e.g., matrix factorization in recommender systems~\cite{zhao2018learning},  
recent TL research has shifted to using large \& deep neural network as classifiers, which has
yielded significantly better accuracy~\cite{ni2018perceive,chen2019icient,hu2018mtnet,yuan2019darec}. 
%In fields like computer vision, transferred super deep neural networks have even achieved human-level classification accuracy~\cite{dodge2017study}.
However, this also brought up new challenges: (1) how to perform efficient transfer learning for resource-limited applications? (2) how to avoid overfitting problems for large neural network models when training examples are scarce in the target domain? To our knowledge, these types of research have not been explored in the existing recommendation literature. In fact, we are even not sure  whether it is possible to learn an effective user representation by \textit{only} using their past behaviors (i.e., no user profiles \& no other item features), and whether such representations can be transferred to improve the
%help improve 
downstream tasks.
% ,analogous to learned visual features from ImageNet~\cite{guo2019spottune,kornblith2019better} or embedding features from text corpus~\cite{radford2018improving,houlsby2019parameter,devlin2018bert}. 

Closely related to this work, ~\cite{ni2018perceive} recently introduced a DUPN model, which represents deep user perception network.  DUPN is also capable of learning  general user representations for multi-task purpose. 
But we find there are several key differences from this work. First, DUPN has to be pre-trained by a multi-task learning objective, i.e., more than one training loss. It showed that the learned user representations performed much worse if there are no auxiliary losses and data.
By contrast, PeterRec is pre-trained by one single loss but can be adapted to multiple domains or tasks. To this end, we define the task in this paper as a multi-domain learning problem~\cite{rebuffi2018efficient}, which distinguishes from the multi-task learning in DUPN. Second, DUPN performs pre-training by relying on many additional features, such as user profiles and item features. It requires expensive human efforts in feature engineering, and it is also unclear whether
the user representation work or not without these features. 
%In fact, we find the way of pre-training in DUPN is very weak since it does not explicitly model the item dependencies in the user sequence. Similar issues have already been analyzed by authors in NextitNet when they compare with Caser~\cite{tang2018caser}. 
%By contrast, PeterRec utilizes
%the NextItNet-style  methods for pretraining, which yields more effective user representation.
% the pre-training method in DUPN only for the fine-tuning phase, while applies completely different pre-training methods. 
Third, DUPN
does not consider efficient transfer learning issue since it only investigates fine-tuning all pre-trained parameters and  the final classification layer.
By contrast,  PeterRec fine-tunes a small fraction of injected parameters, but obtains comparable or better results than fine-tuning all parameters.
% and performs notably better than fine-tuning only the classification layer. 
%The research of parameter sharing based TL across multiple domains has not been studies in the field of recommender systems. 

CoNet~\cite{hu2018conet} is another cross-domain recommendation model using neural networks as the base  model.  To enable knowledge transfer, CoNet jointly trains two objective functions, among which one represents the source network and the other the target. One interesting conclusion was made by the authors of CoNet is that the pre-training and fine-tuning paradigm in their paper does not work well  according to the empirical observations.
%the way of pre-training a multilayer perceptron network on the source domain and then transferring user representations to the target domain for fine-tuning does not yield performance improvements~\cite{hu2018conet}. The negative results in CoNet may come from many aspects, such as the way of pre-training, the expressiveness of their user representations, as well as the quality of the pre-training data.
In fact, neither CoNet nor 
DUPN  provides evidence  that fine-tuning with a pre-trained network performs better than fine-tuning from scratch, which, beyond doubt, is the fundamental  assumption for TL in recommender systems. By contrast, in this paper, we clearly demonstrate that the proposed PeterRec notably improves the accuracy of downstream recommendation tasks by fine-tuning on the pre-trained model relative to training from scratch. 

%Recently, DARec~\cite{yuan2019darec} and MTNet~\cite{hu2018mtnet} were proposed for cross-domain recommendations by transferring user rating patterns and text contents, respectively. Since these work solve the TL problems from very different perspectives, we simply omit detailed reviews of them.

\section{PeterRec}
The training procedure of PeterRec consists of two stages. 
The first stage is learning a high-capacity user representation model on datasets with  plenty of  user sequential user-item interactions.
Then there is a supervised fine-tuning stage, where the pre-trained representation is adapted  to the downstream task with supervised labels. In particular, we attempt to share the majority of  parameters.

%\subsection{Model Architectures of pre-trained Models}
%\subsection{pre-trained Models}
\subsection{Notation}
\begin{figure}[t]
	%\hspace*{-0.1cm}
	\centering
	\small	
	\includegraphics[width=0.49\textwidth]{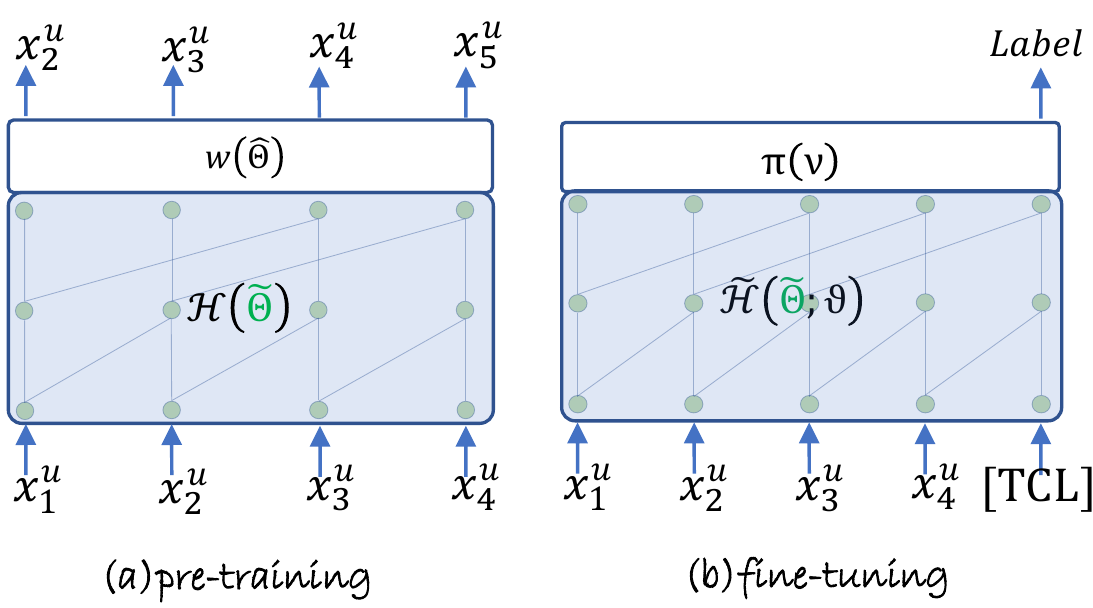}	
	\caption{\small Illustration of parameters in PeterRec.   $x_i^u$denote an itemID in the input sequence of user $u$. 	[TCL] is a special token representing the classification symbol.}
	\label{overviewpeter}
\end{figure}
We begin with some basic notations. 
Suppose that we are given two domains: a source domain $\mathcal{S}$ and target domain $\mathcal{T}$.
%Let  $\mathcal{S}$  and $\mathcal{T}$ be the source and target domain respectively. 
For example, $\mathcal{S}$ can be news or video recommendation where a large number of user interactions are often available, and $\mathcal{T}$
can be a different prediction task  where user labels are usually very limited. In more detail,
a user label in this paper can be an item he prefers in $\mathcal{T}$, an age bracket he belongs to, or the marital status he is in. 
% related to users in $\mathcal{S}$. 
Let $\mathcal{U}$ (of size $|\mathcal{U}|$) be the set of users shared in both domains. Each instance in $\mathcal{S}$ (of size $|\mathcal{S}|$) consists of a userID $u \in \mathcal{U}$, and the unsupervised interaction sequence  $\bm{\mathrm{x}}^u=\{x_1^u, ..., x_n^u \}$ ($x_i^u \in \mathcal{X}$), i.e., $(u,\bm{\mathrm{x}}^u) \in 
% \{\mathcal{U}, \mathcal{X}\}$, 
\mathcal{S}$,
where $x_t^u$ denotes the $t$-th interacted item of $u$ and  $ \mathcal{X}$  (of size $|\mathcal{X}|$)  is the set of items in $\mathcal{S}$.
Correspondingly, each instance in $\mathcal{T}$ (of size $|\mathcal{T}|$) consists of a userID $u$, along with the supervised label $y \in \mathcal{Y}$, i.e., $(u,y) \in \mathcal{T}$. Note if $u$  has $g$ different labels, then there will be $g$ instances for $u$ in $\mathcal{T}$.

We also  show the parameters in the pre-trained and fine-tuned models in Figure~\ref{overviewpeter}.
$\mathcal{H}(\widetilde{\Theta})$ is the pretrained network, where  $\widetilde{\Theta}$ include  parameters of the embedding and convolutional layers;  $w(\hat{\Theta})$ and $\pi(\nu)$ represent  the classification layers for pre-training and fine-tuning,  respectively;  and $\mathcal{\widetilde{H}}(\widetilde{\Theta}; \vartheta)$ is the fine-tuning network with  pre-trained $\widetilde{\Theta}$ and re-learned model patch parameters $\vartheta$. 
$\mathcal{H}(\widetilde{\Theta})$ and $\mathcal{\widetilde{H}}(\widetilde{\Theta}; \vartheta)$  share the same network architecture except the injected model patches (explained later).
%$\Theta$ are learned by pre-training and  shared for all fine-tuning tasks. $\vartheta$ and $\nu$ are domain-specific parameters.

\subsection{User Representation Pre-training}
\label{UserRepL}

{\textbf{Pre-training Objectives.}}
Following NextItNet~\cite{yuan2019simple}, 
%we treat user behavior data with the chronological order as natural sequences, and
we  model the user interaction dependencies in the sequence by a left-to-right chain rule factorization, aka an autoregressive~\cite{bengio2000modeling} method. Mathematically, the joint probability $p(\bm{\mathrm{x}}^u;\Theta)$ of  each user sequence is represented by  the product of the conditional distributions over the items, as shown in Figure~\ref{overviewpeter} (a):
\begin{equation}
\small
\label{l2r}
p(\bm{\mathrm{x}}^u;\Theta)=\prod_{i=1}^{n}p(x_i^u|x_1^u,..,x_{i-1}^u;\Theta)
\end{equation}
where the value $p(x_i^u|x_1^u,...,x_{i-1}^u; \Theta)$ is the probability of the $i$-th interacted item $x_i^u$ conditioned on all its previous interactions $\{x_1^u,...,x_{i-1}^u\}$, $\Theta$ is the parameters of pre-trained model including network  parameters $\widetilde{\Theta}$ and the classification layer parameters $\widehat{\Theta }$. With such a formulation, the  interaction dependencies in $\bm{\mathrm{x}}^u$ can be  explicitly modeled, which is more powerful than existing pre-training approaches (e.g., DUPN) that simply treat the item sequence $\bm{\mathrm{x}}^u$ as common feature vectors.
%by constructing additional supervised training objectives. 
To the best of our knowledge, PeterRec is the first TL model in the recommender system domain that is pre-trained by unsupervised autoregressive approach.

Even though user-item interactions come in the form of sequence data, the sequential dependency may not be strictly held in terms of user preference, particularly for recommendations. This has been verified in~\cite{yuan2020future}, which introduced GRec that estimates the target interaction by considering both past and future interactions.
%Unlike~\cite{yuan2020future}, we are not interested in the generative accuracy of the pre-trained model, but whether using two side contexts helps to improve the 
%the quality and generality of user representations. 
As such, we introduce an alternative pre-training objective by taking account of two-side contexts.
%by
%modifying GRec as a bidirectional representation learning model.
Specifically, we 
randomly mask a certain percentage of items (e.g., 30\%) of $\bm{\mathrm{x}}^u$ by filling in the mask symbols (e.g., ``\_\_'') in the sequence, and then predict the items at these masked position by directly adding a softmax layer on the encoder of GRec. 

Formally, let $\bm{\mathrm{x}}^u_\triangle=\{x_{\triangle_1 }^u,...,x_{\triangle_m }^u\}$ ($1 \leq m < t$) be the masked interactions, and $\tilde{\bm{\mathrm{x}}}^u$ is the sequence of $\bm{\mathrm{x}}^u$ by replacing items in $\bm{\mathrm{x}}^u_\triangle$ with ``\_\_'',
the probability of $p(\bm{\mathrm{x}}^u_\triangle)$ is given as:
\begin{equation}
\small
\label{bid}
p(\bm{\mathrm{x}}^u_\triangle;\Theta)=\prod_{i=1}^{m}p(x^u_{\triangle_i}|{\tilde{\bm{\mathrm{x}}}^u;\Theta})
\end{equation}
To maximize  $p(\bm{\mathrm{x}}^u;\Theta)$ or $p(\bm{\mathrm{x}}^u_\triangle;\Theta)$, it is equivalent to minimize the cross-entropy (CE) loss  $L(\mathcal{S};\Theta)=-\sum_{(u,\bm{\mathrm{x}}^u)\in \mathcal S}\log p(\bm{\mathrm{x}}^u;\Theta)$ and  $G(\mathcal{S};\Theta)=-\sum_{(u,\bm{\mathrm{x}}^u)\in \mathcal S}  \log p(\bm{\mathrm{x}}^u_\triangle;\Theta)$, respectively. 
It is worth mentioning that while similar pre-training objectives have been applied 
in the 
%computer vision~\cite{bibid} and 
NLP~\cite{devlin2018bert} and computer vision~\cite{su2019vl} domains recently,  the effectiveness of them remains completely unknown in recommender systems.
% In addition, both the source and target domains share the same vocabulary in the NLP tasks, which makes transfer much easier than recommender systems where items in the source and target tasks are not sharable.
Hence, in this paper instead of proposing
a new pre-training objective function,  we are primarily interested in showing readers what types of  item recommendation models  can be applied to  user representation learning, and 
how to adapt them for pre-training \& fine-tuning  so as to  bridge the gap between different domains.

{\textbf{Petrained Network Architectures.}} 
The main architecture ingredients of the pre-trained model are  a stack of dilated convolutional (DC)~\cite{yuan2019simple,yuan2020future} layers with exponentially increased dilations and a repeatable pattern, e.g., $\{1, 2,4,8,16, 32,1, 2,4,8,16,32,...,32\}$.
%The pre-trained model of PeterRec is composed of a stack of dilated convolutional (DC) layers with exponentially increased dilation rate, e.g., $\{1, 2,4,8,16, 32,1, 2,\\4,8,16,32\}$.
Every two DC layers are connected by a shortcut connection, called residual block~\cite{he2016deep}. Each DC layer in the block  is followed\footnote{\scriptsize Note there is no standard regarding the orders of the DC layer, the normalization and non-linear layer.  } by a  layer normalization and non-linear activation layer, as illustrated in Figure~\ref{modelpatch}~(a).
%Each residual block is stacked by a DC layers, layer normalization, activation function and a skip connection, as illustrated in Figure~\ref{modelpatch}~(a). 
%To better describe the model patch structure for fine-tuning, we now formally  
Following~\cite{yuan2019simple} and ~\cite{yuan2020future}, the pre-trained network should be built by causal and non-causal CNNs for objective fuctions of Eq.~(\ref{l2r}) and Eq.~(\ref{bid}), respectively.

\begin{figure*}[t]
	%\hspace*{-0.1cm}
	\centering
	\small	
	\includegraphics[width=1.0\textwidth]{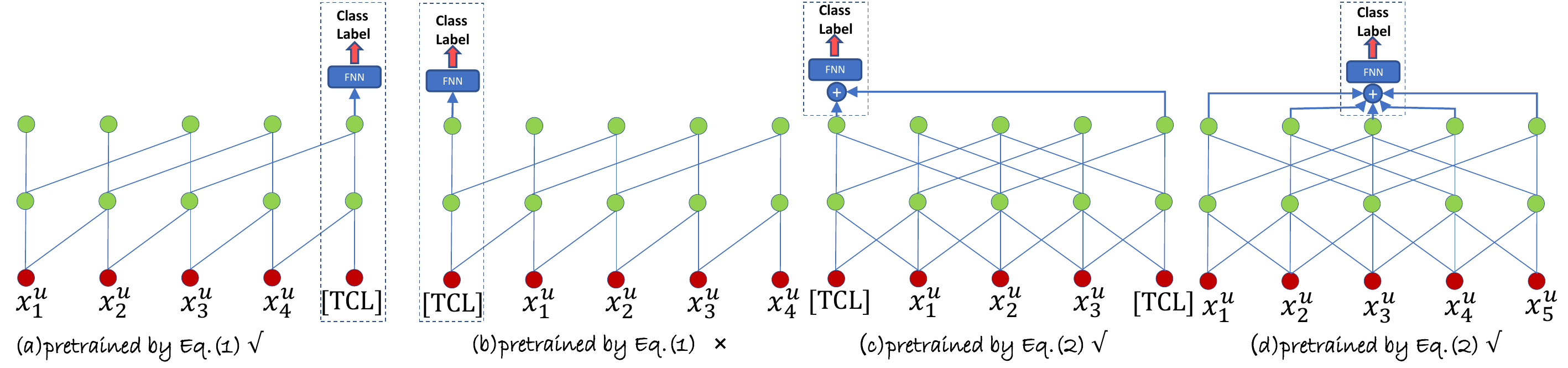}	
	\caption{\small The fine-tuning architecture of PeterRec illustrated with one residual block.  
		Each layer of (green) neurons corresponds to a  DC layer in Figure~\ref{modelpatch}. The normalization layer, ReLU layers and model patches are not depicted here for clearity.
		(a)(b) and (c)(d) are causal and non-causal convolutions, respectively.
		%		The  architecture of fine-tuned PeterRec mainly inherits the pre-trained model  with additional model patches  (see later descriptions in Figure~\ref{modelpatch}) and a classification layer.
		%	inserted in the residual blocks (see later descriptions in Figure~\ref{modelpatch}).
		%		Each layer here contains a DC layer, a model patch,  a normalization layer and an activation function.
		FNN is the feedforward neural network for classification with parameter  $\nu$.
		(a)(c) and (d) are suggested fine-tuning architectures. (b) is not correct since no information can be obtained by causal convolution  if [TCL] is inserted at the beginning. 
		%	Note that the model patches, normalization layers, activation function layers, and the skip connection are omitted here for simplicity.
	}
	\label{finetunearch}
\end{figure*}

Concretely, the residual block with the DC operations is formalized as follows:
\begin{equation}
\small
\begin{aligned}
\mathcal{H}_{DC}(\bm{\mathit{E}})=
\left\{\begin{matrix}
\bm{\mathit{E}}+\mathcal{F}_{cauCNN}(\bm{\mathit{E}})& optimized \ by  \quad Eq.~(1) \\ 
\bm{\mathit{E}}+\mathcal{F}_{non\_cauCNN}(\bm{\mathit{E}})& optimized  \ by  \quad Eq.~(2) 
\end{matrix}\right.
\end{aligned}
\end{equation}
%e operation F + x is performed by a shortcut connection and element-wise addition
where $\bm{\mathit{E}}$  $ \in  \mathbb{R}^{n\times k}$ and $\mathcal{H}_{DC}(\bm{\mathit{E}})\in  \mathbb{R}^{n\times k }$ are the input and output matrices of layers considered,  $k$ is the embedding dimension,  $\bm{\mathit{E}}+\mathcal{F}$ is a shortcut connection by element-wise addition, and
% is the output from the preceding hidden layer, 
$\mathcal{F}_{cauCNN}(\bm{\mathit{E}})$
\& $\mathcal{F}_{non\_cauCNN}(\bm{\mathit{E}})$ are the residual mappings as follows
% , which are learned  by causal and non-causal CNN function, respectively.
\begin{equation}
\label{caus}
\small
\begin{aligned}
\mathcal{F}_{cauCNN}(\bm{\mathit{E}})&=\sigma(\mathcal {LN}_2(\psi_2 (\sigma(\mathcal {LN}_1(\psi_1 (\bm{\mathit{E}})))))) \\
\mathcal{F}_{non\_cauCNN}(\bm{\mathit{E}})&=\sigma(\mathcal {LN}_2(\phi_2 (\sigma(\mathcal {LN}_1(\phi_1 (\bm{\mathit{E}}))))))
\end{aligned}
\end{equation}
where $\psi$ and $\phi$ represent  causal (e.g., Figure~\ref{finetunearch} (a) \& (b) and non-causal (e.g.,  (c) \& (d))  convolution operations, respectively, and the biases  are omitted for shortening notations. $\mathcal {LN}$ and $\sigma$ represent layer normalization~\cite{ba2016layer} and ReLU~\cite{nair2010rectified}, respectively.

\subsection{ User Representations Adapting}
\label{Adapting}
%\subsection{ Supervised Fine-tuning}

After the above pre-training process, we can adapt the learned representations to specific downstream tasks. 
%As opposed to existing works that either treat pre-trained parameters as freezing features or optimize all of them,
%we develop neural network architectures that 
The primary goal here is to develop  a fine-tuning framework that works well in multiple-domain settings by introducing only a small fraction of domain-specific parameters, and attain a high-degree of parameter sharing between domains. Specifically, the architecture of fine-tuning PeterRec contains three components as shown in  Figure~\ref{overviewpeter} (b):  %the pre-trained model by removing its classification layer, 
all except the classification layer of the pre-trained model (parameterized by $\widetilde{\Theta}$),
the new classification layer  (parameterized by $\nu$) for the corresponding downstream task, and the model patches (parameterized by $ \vartheta$)  that are inserted in the pre-trained residual blocks. 
In the following, we first present the overall fine-tuning framework. Then, we describe the details of the grafting patch structure and show how to inject it into the pre-trained model.

%We show the fine-tuning PeterRec in Figure~\ref{finetunearch}, and 
{\textbf{Fine-tuning Framework.}} Let assume that the model patches have  been inserted and initialized in the pre-trained model.
%, which results in the core component of the fine-tuning network. 
The overall  architectures of PeterRec are illustrated in Figure~\ref{finetunearch}.
%PeterRec loads the pre-trained parameters and randomly initializes the model patch modules. 
As a running example, we describe in detail the fine-tuning procedures using the causal CNN network, as shown in (a). For each instance $(u,y)$ in $\mathcal{T}$,
we first add a [TCL] token at the end position of user sequence $u$, and achieve the new input, i.e., $\bm{\mathrm{x}}^u=\{x_1^u, ..., x_n^u, \text{[TCL]} \} $.  
Then, we feed this input sequence to the fine-tuning neural network. By performing a series of causal CNN operations on the embedding of $\bm{\mathrm{x}}^u$, we obtain the last hidden layer matrix. Afterwards, a linear classification layer is placed on top of the  final hidden vector  of the [TCL] token, denoted by $\bm{h}_{n} \in  \mathbb{R}^{ k}$. %$\mathcal{H}^j_{DC}(\bm{\mathit{E}})$  corresponding to the index of `[TCL]', 
Finally, we are able to achieve the scores $\bm{o} \in \mathbb{R}^{|\mathcal{Y}|}$ with respect to all labels in $\mathcal{Y}$, and the probability to predict $y$.
%and $y^\_$, where $y$ and $y^\_$  represent the true and false labels, respectively.
\begin{equation}
\small
\begin{aligned}
\bm{o}&=\bm{h}_n\bm{W}+\bm{b}\\
p(y|u)&=p(y|\bm{\mathrm{x}}^u)=softmax(\bm{o}_y)\\
%&p(y^\_|u)=p(y^\_|\bm{\mathrm{x}}^u)=softmax(\bm{o}_{y^\_})
\end{aligned}
\end{equation}
where $\bm{W} \in  \mathbb{R}^{k \times |\mathcal{Y}|} $ and $\bm{b} \in \mathbb{R}^{|\mathcal{Y}|} $ are the projection matrix and bias term.

In terms of the pre-trained model by non-causal CNNs, PeterRec can simply add [TCL]s at the start and the end positions of $\bm{\mathrm{x}}^u$, as shown in Figure~\ref{finetunearch}~(c),
i.e., $\bm{\mathrm{x}}^u=\{\text{[TCL]}, x_1^u, ..., x_n^u, \text{[TCL]} \}$, and accordingly
%\footnote{\scriptsize [TCL] and the blank token, i.e., ``\_\_'', used in the pre-training phase can be the represented by a same index value.}
\begin{equation}
\small
\begin{aligned}
\bm{o}=(\bm{h}_0+\bm{h}_n) \bm{W}+\bm{b}
\end{aligned}
\end{equation} 
Alternatively, PeterRec can use the sum of all hidden vectors of $\bm{h}$ without adding any [TCL] for both causal and non-causal CNNs, e.g., Figure~\ref{finetunearch}~(d).
\begin{equation}
\small
\begin{aligned}
\!\! \!\!\!\!   \bm{o}=(\sum_{i=1}^{n}\bm{h}_i )\bm{W}+\bm{b}
\end{aligned}
\end{equation} 
Throughout this paper, we will use  Figure~\ref{finetunearch} (a) for causal CNN and (c) for non-causal CNN in our experiments.
%Other feedforward process  during fine-tuning are exactly the same as mentioned above. 

As for the fine-tuning objective functions of PeterRec, we adopt the pairwise ranking loss (BPR)~\cite{rendle2009bpr,lei2020estimation} for
top-$N$ item recommendation task and the CE loss for the  user profile classification tasks.
%\begin{equation}
% \label{lossfunctions}
%\begin{aligned}
% R_{BPR}(\mathcal{T})&=-\sum_{(u,y) \in \mathcal{T}}\log \delta (\bm{o}_y-\bm{o}_{y^\_}) )\\
% R_{CE}(\mathcal{T})&=-\sum_{(u,y) \in \mathcal{T}}\log p(y|u) 
%\end{aligned}
%\end{equation}
\begin{equation}
\label{lossfunctions}
\small
\begin{aligned}
R_{BPR}(\mathcal{T};\tilde{\Theta};\nu;\vartheta)&=-\sum_{(u,y) \in \mathcal{T}}\log \delta (\bm{o}_y-\bm{o}_{y^\_}) \\
R_{CE}(\mathcal{T};\tilde{\Theta};\nu;\vartheta)&=-\sum_{(u,y) \in \mathcal{T}}\log p(y|u) 
\end{aligned}
\end{equation}
%\begin{equation}
%\label{lossfunctions}
%\begin{aligned}
%R_{CE}(\mathcal{T})=-\sum_{(u,y) \in \mathcal{T}}\log p(y|u) 
%\end{aligned}
%\end{equation}
where 
$\delta$ is the logistic sigmoid function, and $y^\_$ is a false label randomly sampled from $\mathcal{Y}\backslash y$ following~\cite{rendle2009bpr}. 
Note that in~\cite{yuan2016lambdafm,yuan2018fbgd}, authors showed that a  properly developed dynamic negative sampler usually performed better than the random one if $|\mathcal{Y}|$ is huge. However, this is beyond the scope of this paper, and we leave it as future investigation.
Eq.(\ref{lossfunctions}) can be then optimized by SGD or  its variants such as Adam~\cite{kingma2014adam}. For each downstream task, PeterRec only updates $\vartheta $ and $\nu$ (including  $\bm{W}$ \& $\bm{b}$) by freezing pre-trained parameters $\widetilde{\Theta}$.
%, given as

%The update equation is given as 
%\begin{equation}
%\label{gradient}
%\small
%\begin{aligned}
%&\vartheta_{new}=\vartheta_{old}- \eta  \frac{\partial R (\widetilde{\Theta};\nu ; \vartheta)}{\partial \vartheta} \\
%&\nu_{new}=\nu_{old}- \eta  \frac{\partial R (\widetilde{\Theta};\nu; \vartheta)}{\partial \nu} 
%\end{aligned}
%\end{equation}
%where $\eta$ is the learning rate. We omit the descriptions of gradient equations $\frac{\partial R (\tilde{\Theta};\nu ; \vartheta)}{\partial \vartheta}$ \& $\frac{\partial R (\tilde{\Theta};\nu; \vartheta)}{\partial \nu}$, which can be automatically calculated by chain rule in deep learning libraries, such as Tensorflow\footnote{\scriptsize\url{https://www.tensorflow.org}}.
%There are several different ways to perform fine-tuning, as shown in 

\begin{figure*}[h]
	%\hspace*{-0.1cm}
	\centering
	\small	
	%		\vspace{-0.05in}
	\includegraphics[width=1.0\textwidth]{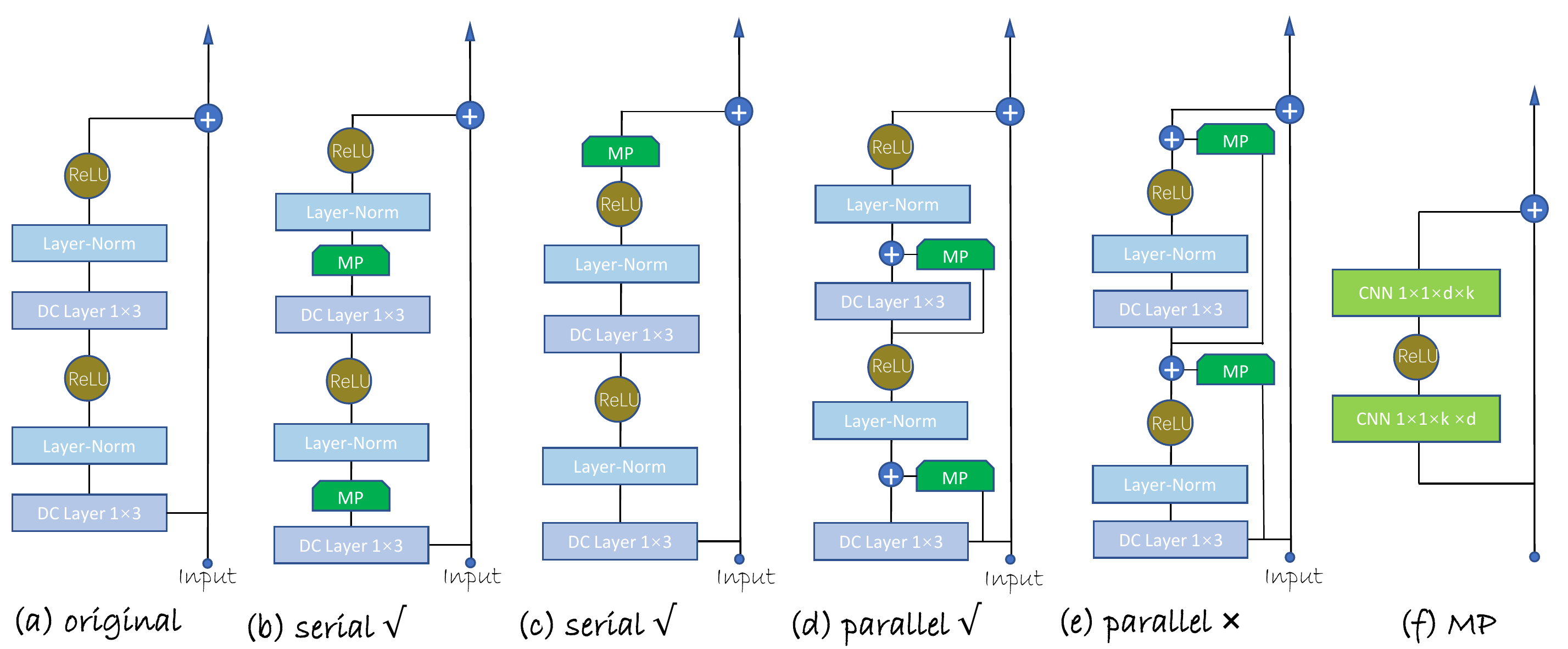}	
	\caption{\small Model patch (MP)  and insertion methods. (a) is the original pre-trained residual block; (b) (c)
		(d)  (e) are the fine-tuned residual blocks with inserted MPs; and (f) is the MP block. $+$ is the addition operation. $1\times 3$ is the kernel size of dilated convolutional layer. }
	\label{modelpatch}
\end{figure*}

\textbf{Model Patch Structure.}
The model patch is a parametric neural network, which adapts the pre-trained DC residual blocks to corresponding tasks, similar to grafting for plants. Our work is motivated and inspired by recent learning-to-learn approaches in~\cite{denil2013predicting,rebuffi2018efficient,mudrakarta2018k} which show that it is possible to predict up to 95\%  of the model parameters given only  the remaining 5\%. Instead of predicting parameters,  we aim to demonstrate how to modify the pre-trained network to obtain better accuracy in related but very different tasks by training only few parameters. 

The structure of the model patch neural network is shown in Figure~\ref{modelpatch}~(f). We construct it using a simple residual network (ResNet) architecture with two $1\times 1$ convolutional layers considering its strong learning capacity in the literature~\cite{ he2016deep}. To minimize the number of parameters, we propose a bottlenet architecture~\cite{yuan2020future}. Specifically, the model patch consists of a projection-down layer, an activation function, a projection-up layer, and a shortcut connection, where  the projection-down layer projects the original $k$ dimensional channels to $d$ ($d \ll k$, e.g., $k=8d$)\footnote{\scriptsize Throughout this paper, we used $k=8d$, though we did not find  that $k=16d$ degraded accuracy.} by  $1\times 1 \times k \times d $ convolutional operations $\phi_{down}$, and the projection-up layer is to project it back to its original dimension by 
$1\times 1 \times d\times k $ convolutional operations $\phi_{up}$. Formally, given its input tensor $\bm{\mathit{\widetilde{E}}} $, the output of the model patch can be expressed as:

\begin{equation}
\begin{aligned}
\mathcal{H}_{MP}(\bm{\mathit{\widetilde{E}}})=\bm{\mathit{\widetilde{E}}}+\phi_{up}(\sigma (\phi_{down}(\bm{\mathit{\widetilde{E}}}))) 
\end{aligned}
\end{equation}
Suppose that the kernel size of the original dilated convolutions is $1\times3$, the total number of the parameters of each DC layer is $3* k^2=192 d^2$, while the number of the patched neural network is $2*k*f=16 d^2$, which is less than 10\% parameters of the original DC network. Note that parameters of biases and layer normalization are not taken into account since  the numbers are  much smaller than that of the DC layer.
Note that using other similar structures to construct the model patch may also perform well, such as in~\cite{xie2017aggregated}, but it generally needs to meet three requirements: (1) to have a much smaller scale compared with the original convolutional neural network; (2) to guarantee that the pre-trained parameters are left unchanged during fine-tuning; and (3)  to attain good accuracy. 

{\textbf{Insertion Methods.}}
Having developed the model patch architecture, the next question is how to inject it into the current DC block. 
We introduce two ways for  insertion, namely serial \& parallel mode patches, as shown in  Figure~\ref{modelpatch} (b) (c) (d) and~(e). 

First, we give the formal mathematical formulations of the fine-tuning block (by using causal CNNs as an example) as follows:
\begin{equation}
\small
\begin{aligned}
\mathcal{\widetilde{H}}_{DC}(\bm{\mathit{E}})=\bm{\mathit{E}}+\mathcal{\widetilde{F}}_{cauCNN}(\bm{\mathit{E}}) 
\end{aligned}
\end{equation}
where $\mathcal{\widetilde{F}}_{cauCNN}$, short for $\mathcal{\widetilde{F}}$ below, is 

\begin{equation}
\label{caus}
\small
\begin{aligned}
\mathcal{\widetilde{F}}=
\left\{\begin{matrix}
\sigma(\mathcal {LN}_2(\mathcal{H}_{MP2}(\psi_2 (\sigma(\mathcal {LN}_1(\mathcal{H}_{MP1}(\psi_1 (\bm{\mathit{E}}))))))) )& Figure~\ref{modelpatch}~(b)  \\ 
\mathcal{H}_{MP}(\sigma(\mathcal {LN}_2(\psi_2 (\sigma(\mathcal {LN}_1(\psi_1 (\bm{\mathit{E}})))))) )& Figure~\ref{modelpatch}~(c)  \\
\sigma(\mathcal {LN}_2(\mathcal{H}_{MP2}(\bm{\mathit{\widetilde{E}}})+ \psi_2(\bm{\mathit{\widetilde{E}}})),   & Figure~\ref{modelpatch}~(d)  \\
where \ \bm{\mathit{\widetilde{E}}}=\sigma(\mathcal {LN}_1(\mathcal{H}_{MP1}(\bm{\mathit{E}})+\psi_1 (\bm{\mathit{E}})))))&\\
\mathcal{H}_{MP2}(\bm{\mathit{\widetilde{E}}})+\sigma(\mathcal {LN}_2(\psi_2 (\bm{\mathit{\widetilde{E}}}))),  & Figure~\ref{modelpatch}~(e)\\
where \  \bm{\mathit{\widetilde{E}}} =\mathcal{H}_{MP1}(\bm{\mathit{E}})+\sigma(\mathcal {LN}_1(\psi_1 (\bm{\mathit{E}}))) &
%\sigma(\mathcal {LN}_1(\mathcal{H}_{MP1}(\bm{\mathit{E}})+\psi_1 (\bm{\mathit{E}})))))
\end{matrix}\right.
\end{aligned}
\end{equation}

In fact, as shown in Figure~\ref{modelpatch}, we only suggest architectures of (b) (c) and (d) as (e) usually converges and performs significantly worse as evidenced and explained in Section~\ref{rq4}.  Here, we give several empirical principles on how to insert this model.

\begin{itemize}
	\item For the serial insertion, the inserted positions are very flexible so that one can inject the grafting patches either before or after layer normalization, as shown in (b) and (c).
	\item For the serial insertion, the number of patches for each DC residual block is very flexible so that one can inject either one or two patches. It gives almost the same results if $k$ in (c) is two times larger than that in (b).
	\item  For parallel insertion, PeterRec is sensitive to the inserted positions, as shown in (d) and (e). Specifically,
	the model patch that is injected before layer normalization (i.e., (d)) performs better than that  
	between layer normalization and activation function, which  performs largely better than that 
	after activation function (i.e., (e) ) .
	\item For parallel insertion, 
	%	it has a strict requirement for  the number of model patches of each DC residual block.
	%	 Empirically, 
	PeterRec with two  patches inserted in the DC block usually  performs slightly better than that with only one patch. 
	%	 despite that  $k$ in the one model patch is two times larger. 
\end{itemize}

In practice, both the serial and parallel insertions with a proper design can yield comparable results with fine-tuning the entire model.
%, which, however, only need to tune a small set of domain-specific parameters.
Let us give a quantitative analysis regarding the number of tuned parameters. 
Assuming that PeterRec utilizes  500,000 items from a source domain, 1024 embedding \& hidden dimensions,  20 residual blocks (i.e., 40 layers), and 1000 class labels  to be predicted in the target domain, the overall parameters are $500,000 * 1024+1024*1024*3 \ (\text{here 3  is the kernel size} ) \ *40+1024*1000 \approx 639$ million,  the number of tuned parameters for $\vartheta$ and $\nu$ is  $2*1024*1024/8*40 \approx 10$ million and $1024*1000 \approx 1$ million, respectively, which in total takes less than 1.7\% of the number of all parameters. Note that (1) the number of parameters $\nu$  can never be shared due to the difference of the output space in the target task, and it  depends on the specific downstream task. It may be  large if the task  is an item recommendation task and may be very small if the task is user modeling (E.g., for gender estimation, it is $1024*2 =2048$);  (2) Though there are several ways to compress the input embedding and output classification layers, which can lead to really large compression rates~\cite{sun2020generic,anderson2020superbloom}, we do not describe them in detail as this is clearly beyond the scope of our paper.

%Serial mode patch: we show the serial insertion method in Figure~\ref{modelpatch}~(b) \& (c). As seen, it is flexible for the serial insertion since the patches can be inserted in different positions in the residual block, and using one model patch for each block can work as effectively as using two.

%all timesteps in the groud truth $\textbf{x}$ is known. 
% For the generating process, the prediction are sequential: after each item is predicted, it is fed back to the network to predict the next one. 
%But also note that both the training and evaluation of the distributions  over the item values are computed in parallel. 
\section{Experiments}
\label{EXPERIMENTS}

%As the key contribution of this work is to develop a new adversarial learning method APR for personalized ranking, we aim to answer the following research questions via experiments. 
%As the key contribution of this work is to improve the existing left-to-right style learning algorithms for SRS, we
%evaluate the proposed approaches on four real-world datasets and conduct detailed ablation studies  to answer the following research questions:
In our experiments, we answer the following research questions:
\begin{enumerate}
	\item \textbf{RQ1:} 
	% Whether the proposed PeterRec work or not for the  downstream tasks, or
	Is the self-supervised learned user representation really helpful for the downstream tasks? To our best knowledge, as a fundamental research question for transfer learning in the  recommender system domain, this has never been verified before. 
	\item \textbf{RQ2:}  How does PeterRec perform with the proposed model patch compared with fine-tuning the last layer and  the entire model?
	\item \textbf{RQ3:}  What types of user profiles can be estimated by PeterRec?  Does PeterRec work well  when users are cold or new in the target domain.
	\item \textbf{RQ4:} Are there any other interesting insights we can draw by the ablation analysis of PeterRec?
	% E.g., which type of pre-trained network helps to yield  better fine-tuning accuracy? 
	% Does a pre-trained model that performs better on the source task also perform better on the target tasks? 
\end{enumerate}
\subsection{Experimental Settings}
%\subsection{Datasets and Experiment Setup}
%\subsubsection{\textbf{Datasets}} 

%
%\begin{table} 
%	\centering
%	\caption{\small Number of instances.  Each instance in $\mathcal{S}$ and $\mathcal{T}$ represents $(u,\bm{\mathrm{x}}^u)$ and $(u,y)$ pairs, respectively. The number of source items $|\mathcal{X}|$=$ 191K$, $645K$, $645K$, $645K$, $645K$ ($K=1000$), and the number of target  labels $|\mathcal{Y}|$=$20K$,$ 17K$, $2$, $8$, $6$ for the five dataset from left to right in the below table.}
%	\small
%	\label{datasets}
%	\setlength{\tabcolsep}{1.6mm}
%	\begin{threeparttable}				
%		\begin{tabular}{c|c|c|c|c|c}
%			\toprule
%			\small Domain &  \small ColdRec-1& \small ColdRec-2 &  \small GenEst&  \small AgeEst&  \small LifeEst\\
%			\midrule
%			$\mathcal{S}$        &1,649,095 &1,551,881 &-& -&-\\ 	
%			\midrule
%			$\mathcal{T}$     & 3,798,114 & 2,947,688 & 1,548,844& 1,551,357& 1,075,010\\ 				
%			\bottomrule
%		\end{tabular}
%	\end{threeparttable}
%	%	\vspace{-0.1in}
%\end{table}

\begin{table} 
	\centering
	\caption{\small Number of instances.  Each instance in $\mathcal{S}$ and $\mathcal{T}$ represents $(u,\bm{\mathrm{x}}^u)$ and $(u,y)$ pairs, respectively. The number of source items $|\mathcal{X}|$=$ 191K$, $645K$, $645K$, $645K$, $645K$ ($K=1000$), and the number of target  labels $|\mathcal{Y}|$=$20K$, $ 17K$, $2$, $8$, $6$ for the five dataset from left to right in the below table. $M=1000K$.}
	\small
	\label{datasets}
	\setlength{\tabcolsep}{2.1mm}
	\begin{threeparttable}				
		\begin{tabular}{c|c|c|c|c|c}
			\toprule
			\small Domain &  \small ColdRec-1& \small ColdRec-2 &  \small GenEst&  \small AgeEst&  \small LifeEst\\
			\midrule
			$\mathcal{S}$        &$1.65M$ &$1.55M$ &-& -&-\\ 	
			\midrule
			$\mathcal{T}$     & $3.80M$ & $2.95M$ & $1.55M$& $1.55M$&$ 1.08M$\\ 				
			\bottomrule
		\end{tabular}
	\end{threeparttable}
	%	\vspace{-0.1in}
\end{table}

%\begin{table}[thpb]
%	\caption{Inference speedup. The values denote multiples.  M is slot number.}
%	\label{tab:time_reduce}
%	\begin{tabular}{c|cccccc|c}
%		\toprule
%		M&2&3&4&6&9&12&Average  \\
%		\midrule
%		weishi&1.51&1.52&--&--&--&--&1.515  \\
%		ml-10M&6.28&--&6.71&6.17&4.68&--&5.96  \\
%		ml-latest&11.35&--&12.16&10.51&8.28&8.03&10.07\\
%		\bottomrule
%	\end{tabular}
%\end{table}

{\textbf{Datasets.}} 
We conduct experiments on several large-scale industrial datasets collected by the Platform and Content Group of Tencent\footnote{\scriptsize\url{https://www.tencent.com/en-us/}}. 
%Both source codes and datasets after data masking will be released to advance the transfer learning research in recommender systems.
%, and we hope PeterRec will serve as a benchmark for future research.

\textbf{1. ColdRec-1 dataset}: This contains both source and target datasets. The source dataset is the news recommendation data collected from QQ Browser\footnote{\scriptsize\url{https://browser.qq.com}} recommender system from 19th to 21st, June, 2019. Each interaction denotes a positive feedback (e.g., full-play or thumb-up) by a user at certain time. For each user, we construct the sequence using his recent 50 watching interactions by the chronological order. For users that have less than 50 interactions, we simply pad with zero in the beginning of the sequence following common practice~\cite{yuan2019simple}. The target dataset is collected from Kandian\footnote{\scriptsize\url{https://sdi.3g.qq.com/v/2019111020060911550}} recommender system in the same month where an interaction can be  be a piece of news, a video or an advertisement. All users in Kandian are cold with at most 3 interactions (i.e., $g \leq 3$) and half of them have only one interaction. All users  in the target dataset  have corresponding records in the source dataset.  
%Note QQ Browser and  Kandian are completely different products with own items and users, though there are many users shared in both scenarios.

\textbf{2. ColdRec-2 dataset}: It has similar characteristics with ColdRec-1. The source dataset contains 
recent 100 watching interactions of each user, including both news and videos. The users in the target dataset have at most 5 interactions (i.e., $g \leq 5$).
%, and  half of them have only one interaction.

\textbf{3. GenEst dataset}: It has only a target dataset since all users are from the source dataset of ColdRec-2.
%That is, the pre-trained model of ColdRec-2  is reused.
Each instance  in GenEst is a user and his  gender (male or female) label ($g=1$) obtained by the registration information. 

\textbf{4. AgeEst dataset}: Similar to GenEst, each instance  in AgeEst is a user and his age bracket label ($g=1$) --- one class represents 10 years.

\textbf{5. LifeEst dataset}:  Similar to GenEst, each instance in LifeEst is a user and  his life status label ($g=1$), e.g., single, married, pregnancy or parenting. 

Table \ref{datasets} summarizes other statistics of evaluated datasets.

\noindent{\textbf{Evaluation Protocols.}} 
To evaluate the performance of PeterRec in the downstream tasks, we
randomly split the target dataset into training (70\%), validation (3\%) and testing  (27\%)  sets. We use  two popular top-$5$ metrics ---  MRR@$5$ (Mean Reciprocal Rank)~\cite{yuan2018fbgd} and HR@$5$ (Hit Ratio)~\cite{he2017neural, he2020lightgcn} --- for  the cold-start recommendation datasets (i.e. ColdRecs), 
and the classification accuracy (denoted by Acc, where Acc = number of correct predictions$/$total number of predictions) for the other three  datasets. Note that to speed up the experiments of item recommendation tasks, we follow the common strategy in~\cite{he2017neural} by randomly sampling 99 negative examples for the true example, and evaluate top-$5$  accuracy among the 100 items.

\begin{figure*}
	\small
	\centering     %%% not \center
	\subfigure[\scriptsize Cold-Rec1 (one epoch:  $50*b$  ) ]{\label{yahoo-alpha}\includegraphics[width=0.19\textwidth]{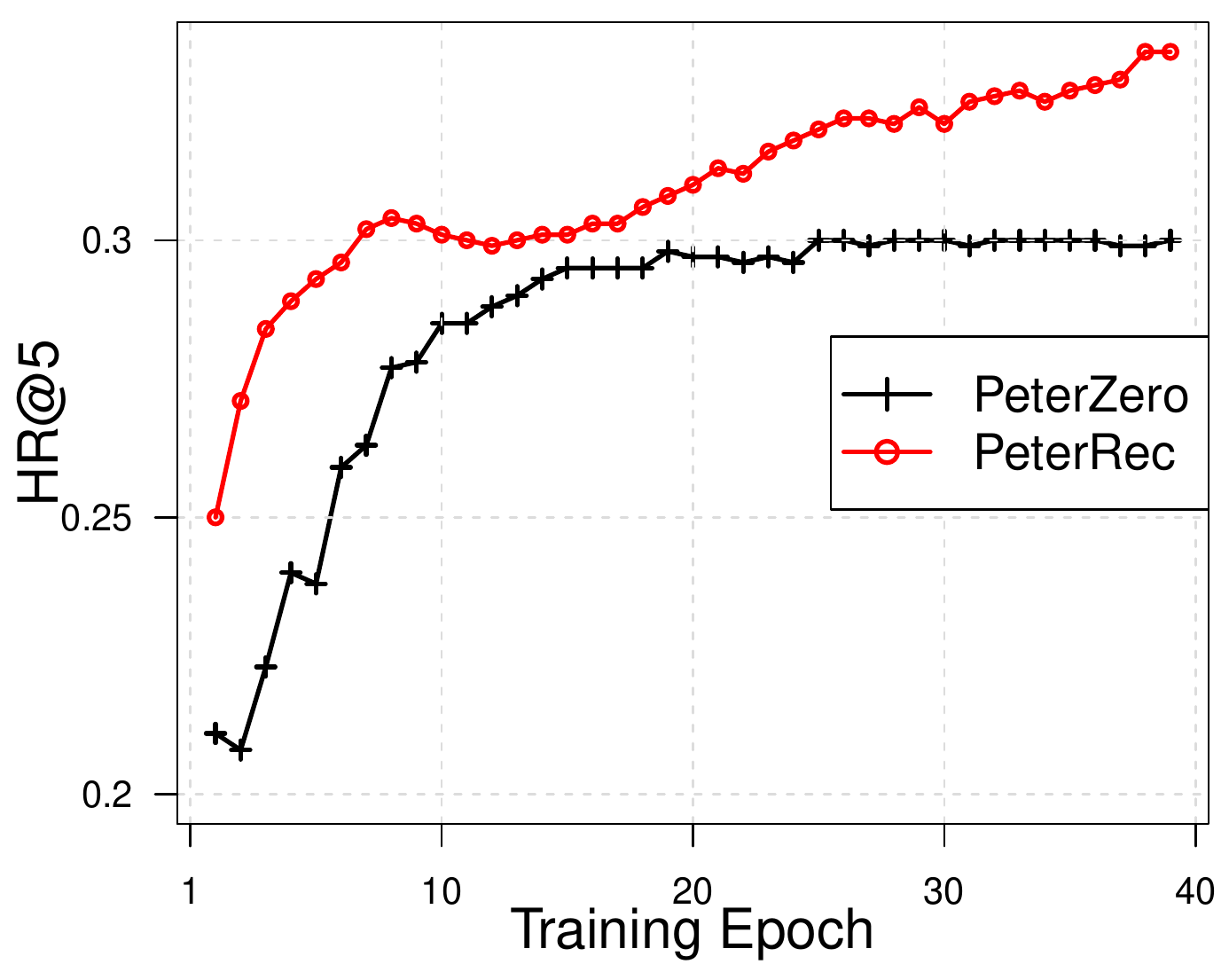}}
	\subfigure[\scriptsize Cold-Rec2  (one epoch:  $50*b$) ]{\label{yahoo-alpha}\includegraphics[width=0.19\textwidth]{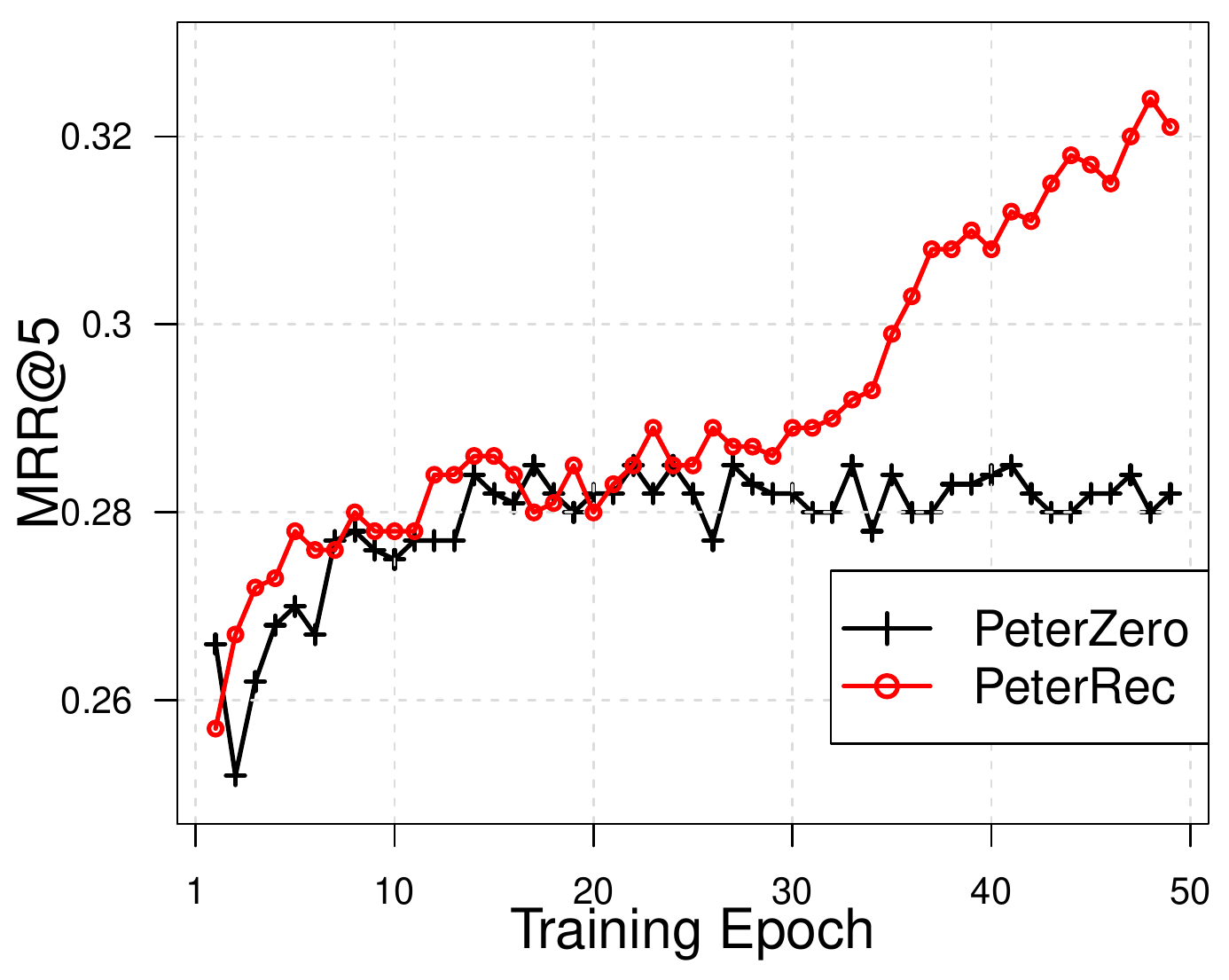}}
	\subfigure[\scriptsize GenEst  (one epoch : $5*b$) ]{\label{yahoo-alpha}\includegraphics[width=0.19\textwidth]{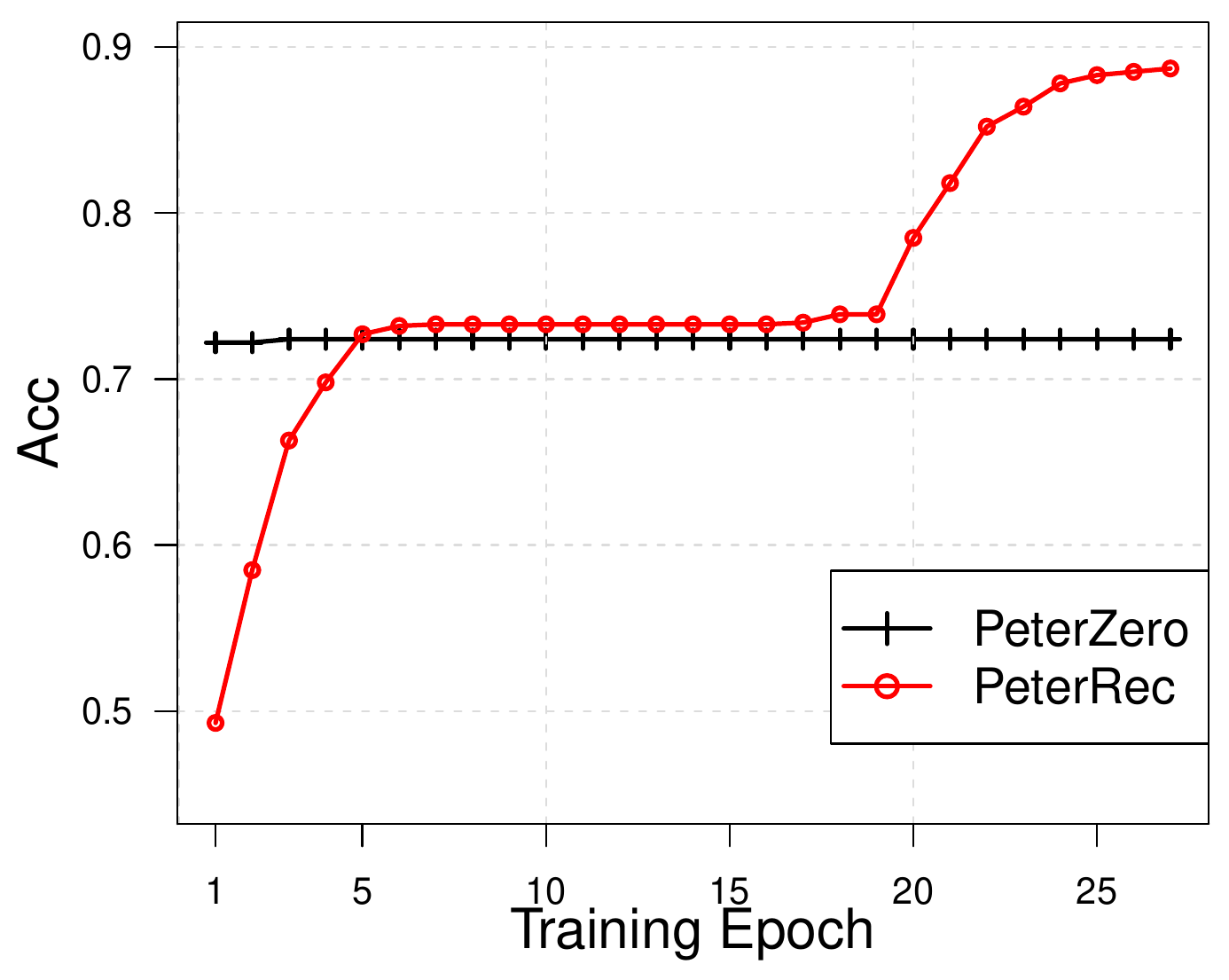}}
	\subfigure[\scriptsize AgeEst  (one epoch:  $50*b$) ]{\label{yahoo-alphazero}\includegraphics[width=0.19\textwidth]{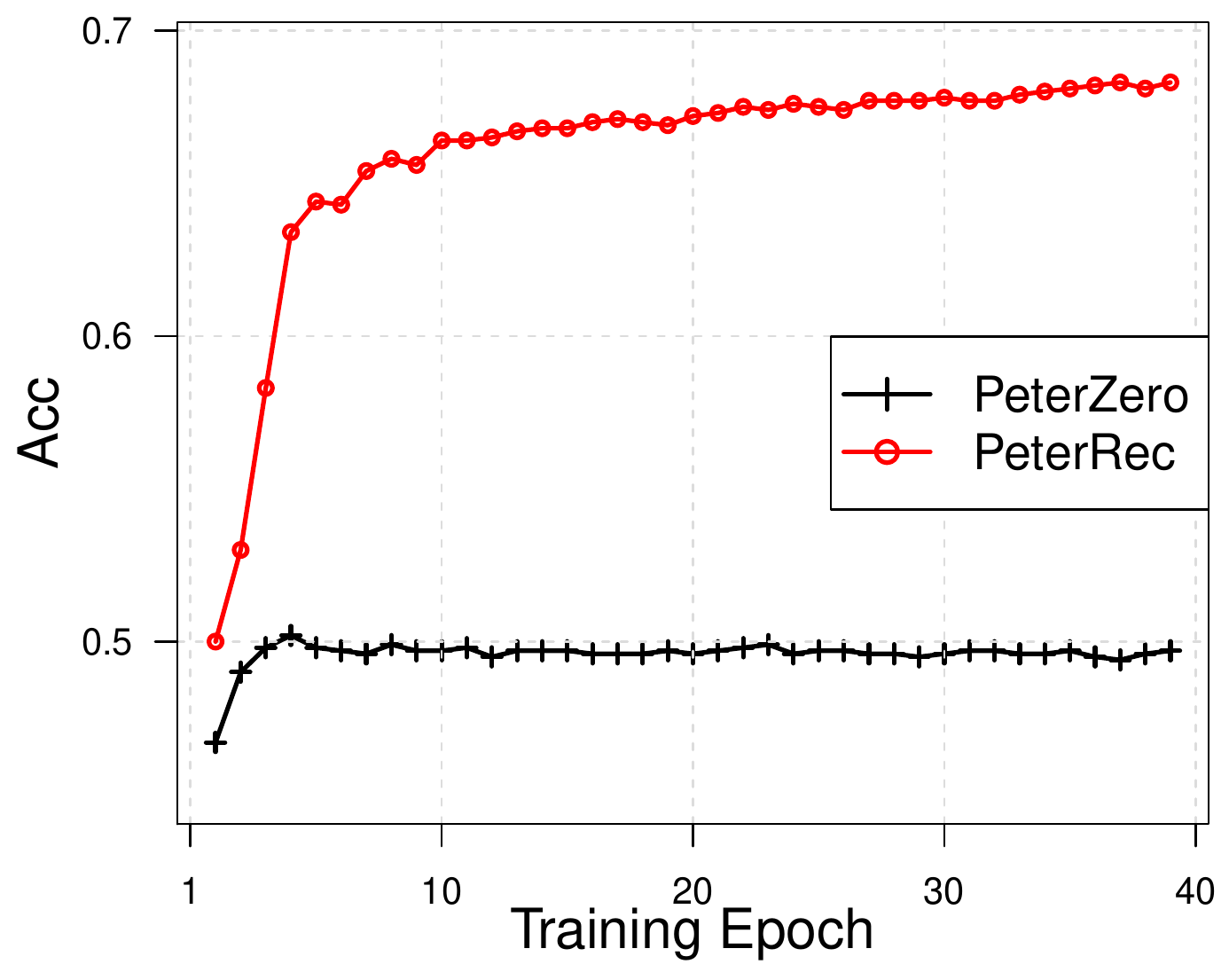}}
	\subfigure[\scriptsize LifeEst ( one epoch:  $5*b$ ) ]{\label{yahoo-alpha}\includegraphics[width=0.2\textwidth]{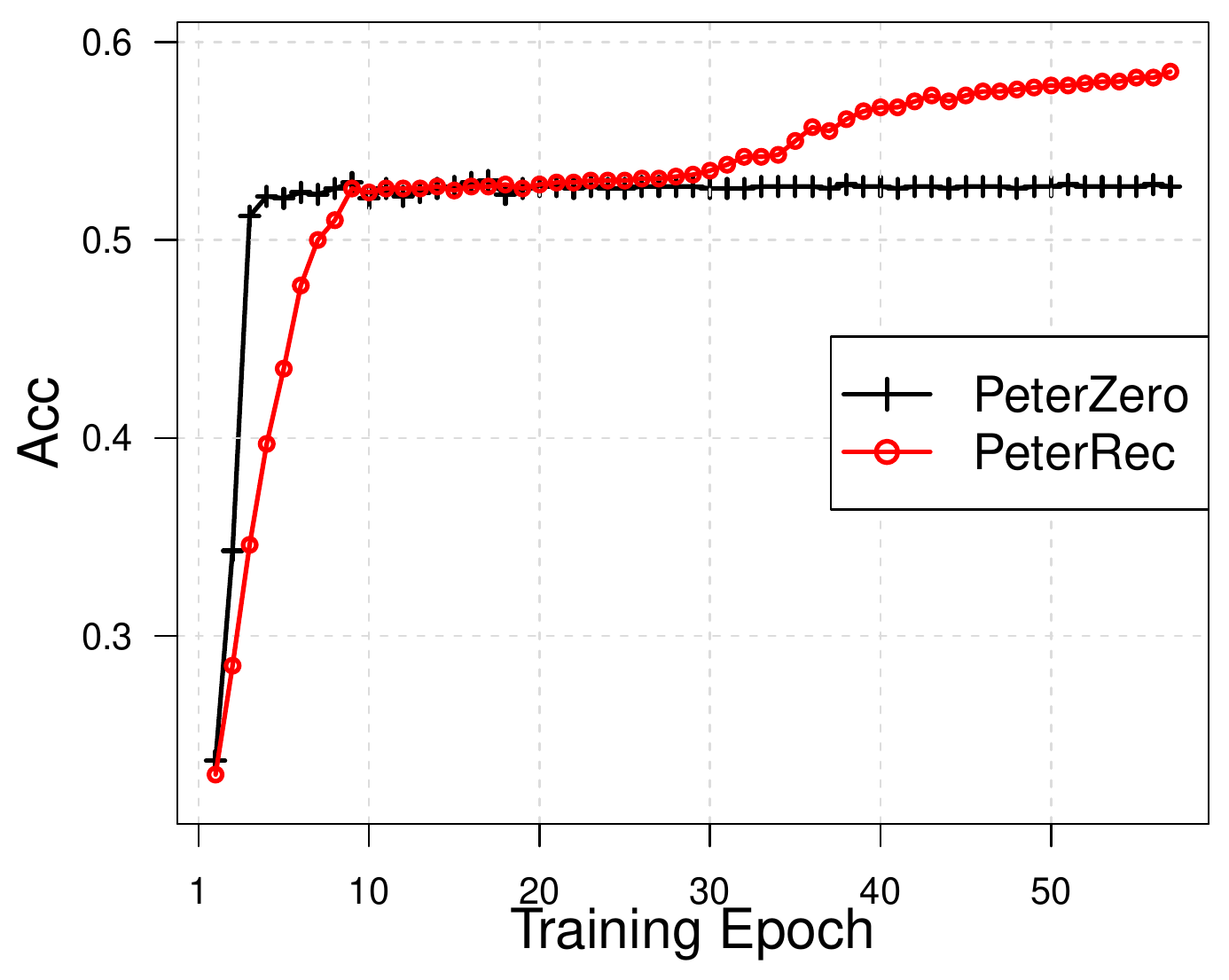}}
	\caption{\small Impact of pre-training  --- PeterRec (not converged) vs. PeterZero (fully converged)  with the causal CNN. $b$ is batch size. Note that since PeterZero converges much faster (and worse) in the first several epoches, here we only show the results of  PeterRec for these beginning epoches for a better comparison. The converged results are given in Table~\ref{finezeroandfineall}. } 
	\label{impactofpre-training}
\end{figure*}
%MRR and NDCG take the rank of the item into  account, which is important in settings where the order of  recommendations matters, while HR@$N$ does not consider the actual rank of the item as long as it is amongst the  top-$N$.
%For saving spaces, we have omitted the detailed formulas. 
\noindent{\textbf{Compared Methods.}} 
We  compare  PeterRec with the following baselines  to answer the proposed research questions. 

\begin{itemize}
	\item To answer \textbf{RQ1}, we compare PeterRec in two cases: well-pre-trained and no-pre-trained settings. We refer to PeterRec with randomly initialized weights as PeterZero.
	\item To answer \textbf{RQ2},  we  compare PeterRec with three  baselines which initialize their weights using pre-trained parameters: (1) fine-tuning only the linear classification layer that is designed for the target task, i.e.,  treating the pre-trained model as a feature extractor, referred to  as FineCLS;  (2) fine-tuning the last CNN layer, along with the linear classification layer of the target task,  referred to  as  FineLast; (3)  fine-tuning  all parameters, including both the pre-trained component (again, excluding its softmax layer) and the linear classication layer for the new task, referred to  as FineAll.
	\item To answer \textbf{RQ3}, we compare PeterRec with an intuitive baseline,  which performs classification based on the largest number of labels in $\mathcal{T}$, referred to as LabelCS.  
	%	It is also known as the most-popular \cite{yuan2017boostfm,yuan2016lambdafm} (MP) method, a widely used baseline in  recommender systems.
	For cold-start user recommendation, we compare it with two powerful baseline NeuFM~\cite{hexiagnan2017neuralf} and DeepFM~\cite{guo2017deepfm}.  
	%	 In addition, we compare another two powerful baseline NeuFM~\cite{he2017neural} and Youtube DNN model (short for YouDNN)~\cite{covington2016deep}. 
	For a fair comparison, we slightly change NeuFM and DeepFM by treating interacted items in $\mathcal{S}$  as features and  target items as softmax labels, which  has no negative effect on the performance~\cite{rendle2020neural}.
	%	The interacted items in $\mathcal{S}$ are treated as features and target items as labels for the two models, which can  be seen as a feature-based transfer learning. 
	In adition, we also present a multi-task learning (referred to as MTL) baseline  by adapting DUPN~\cite{ni2018perceive} to our dataset, which jointly learns the objective functions of both source and target domains instead of using the two-stage pre-training and fine-tuning schemes of PeterRec.  
	%	Note that PeterRec is not directly comparable with DUPN since (1) its training  process relies on the a large number of item and user behavior (e.g., click, bookmark and purchase, etc.) features; (2) for each user, it requires his other preference labels (e.g., conversion rate, price or fashion styles)  to perform  multi-task learning.
	
	%	In fact, FineZero (FineAll without pre-training) can also work as a  competitive baseline since it has similar expressivity\footnote{\scriptsize We refer to the interested readers to see~\cite{lee2017ability} for a detailed analysis.}  (by stacked convolutional layers) to mimic various embedding or neural network   (e.g., YouTube DNN~\cite{covington2016deep}, variants of Factorization Machines~\cite{le2012factorization, he2017neural}, Google Wide \& Deep~\cite{hexiagnan2017neuralf}, etc.) models that use 
	%	the interacted items in $\mathcal{S}$ as general  features for the classification (or regression) model.
	\item To answer \textbf{RQ4}, we  compare PeterRec by using different settings, e.g., using caus\underline{al} and n\underline{on}-causal CNNs,  referred to as PeterRecal and PeterRecon, respectively.
\end{itemize}

\begin{table} 
	\centering
	\caption{\small Impacts of pre-training --- FineZero vs. FineAll (with the causal CNN architectures).  Without special mention, in the following we only report ColdRec-1 with HR@5  and  ColdRec-2 with MRR@5 for demonstration.  
		%Unlike Figure~\ref{impactofpre-training}, all results are reported when the models are fully converged. 
	}
	\small
	%	\vspace{-0.05in}
	\label{finezeroandfineall}
	\setlength{\tabcolsep}{2.2mm}
	\begin{threeparttable}				
		\begin{tabular}{c|c|c|c|c|c}
			\toprule
			\small Model &  \small ColdRec-1& \small ColdRec-2 &  \small GenEst&  \small AgeEst&  \small LifeEst\\
			\midrule
			%			FineZero        &0.304 &0.290  &0.893& 0.703& 0.596\\ 	
			FineZero        &0.304 &0.290  &0.900& 0.703& 0.596\\ 	
			\midrule
			FineAll       &0.349  & 0.333  & 0.903& 0.710 &  0.610\\ 	
			\midrule
			PeterRec       &0.348  & 0.334  & 0.903& 0.708&  0.610\\ 								
			\bottomrule
		\end{tabular}
		%		\scriptsize \emph{MUSIC\_M5} denotes \emph{MUSIC\_M} with maximum session size of $5$. The same applies to \emph{MUSIC\_L}. `M' denotes 1 million. 
		%    		All models are evaluated in above data sets, although our generative model yields  almost the same prediction accuracy ($\pm 1\%$) without using the manually created sub-sessions.
	\end{threeparttable}
	%		\vspace{-0.1in}
\end{table}
\begin{table} 
	\centering
	\caption{\small Performance comparison (with the non-causal CNN architectures). 
		The number of fine-tuned parameters ($\vartheta $ and $\nu$) of PeterRec accounts for 9.4\%, 2.7\%, 0.16\%, 0.16\%, 0.16\% of FineAll
		on the five datasets from left to right.
	}
	\small
	%	\vspace{-0.05in}
	\label{performcmp}
	\setlength{\tabcolsep}{2.2mm}
	\begin{threeparttable}				
		\begin{tabular}{c|c|c|c|c|c}
			\toprule
			\small Model &  \small ColdRec-1& \small ColdRec-2 &  \small GenEst&  \small AgeEst&  \small LifeEst\\
			\midrule
			%			Source $\mathcal{S}$        &1,649,095 &1,472,428 & & 10& 10\\ 	
			FineCLS        &0.295 &0.293  &0.900& 0.679& 0.606\\ 	
			\midrule
			FineLast       &0.330  & 0.310  & 0.902& 0.682 &  0.608\\ 	
			\midrule
			FineAll   & 0.352& 0.338  & 0.905& 0.714& 0.615\\ 	
			\midrule
			PeterRec& 0.351& 0.339 & 0.906& 0.714& 0.615\\ 	 							
			\bottomrule
		\end{tabular}
		%		\scriptsize \emph{MUSIC\_M5} denotes \emph{MUSIC\_M} with maximum session size of $5$. The same applies to \emph{MUSIC\_L}. `M' denotes 1 million. 
		%    		All models are evaluated in above data sets, although our generative model yields  almost the same prediction accuracy ($\pm 1\%$) without using the manually created sub-sessions.
	\end{threeparttable}
	%	\vspace{-0.1in}
\end{table}
%

%Note in this work, we have simply omitted the comparison with the typical general recommendation models, such as BPRMF~\cite{rendle2009bpr} and NeuMF~\cite{he2017neural}, which generate recommendations based on explicit user embeddings. This kind of models does not work in an extreme cold setting, such as our target task, where half of the users have only one interaction in $\mathcal{T}$.\footnote{\scriptsize If we use these users for training, then they will not be evaluated, and if we use them for evaluation, then they will not be trained.} 
% augmentation methods with two state-of-the-art left-to-right  style CNN baselines, namely, Caser \cite{tang2018caser} and NextItNet \cite{yuan2019simple}. Specifically, Caser is a typical session-based recommendation model based on the left-to-right data augmentation, while NextItNet is a typical left-to-right style model-level augmentation approach. Since both Caser and NextItNet are convolutional neural network (CNN) models, our proposed methods are also learned by CNN for comparison purpose.  
%
%Note that the comparisons to other well-known recommendation models, such as  Bayesian personalized ranking (BPR)~\cite{rendle2009bpr}, Markov chain based models FMC \& FPMC~\cite{rendle2010factorizing} or GRURec~\cite{hidasi2015session,tan2016improved} are simply omitted since they significantly underperform  either Caser or NextItNet in existing lituerature \cite{tang2018caser, yuan2019simple, kang2018self,xiao2019hierarchical}.

\noindent{\textbf{Hyper-parameter Details.}}
All models were trained on GPUs (Tesla P40) using Tensorflow.  All reported results use an embedding \& hidden dimension  of $k$=256. 
%Results for $k$=128, 512 show similar conclusions but are omitted for saving space.  
The learning rates for Adam~\cite{kingma2014adam} with $\eta$ = 0.001 to 0.0001 show consistent  trends.
%Models with $\eta$ =  0.001  usually converge a little bit faster but a little bit  worse than $\eta$ =  0.0001. 
For fair comparison, we use $\eta$ =  0.001 for all compared models on the first two datasets and $\eta$ =  0.0001 on the  other three datasets.
All models including causal and non-causal CNNs use dilation $\{1,2,4,8,1,2,4,8,1,2,4,8,1,2,4,8\}$ (16 layers or 8 residual blocks) following NextItNet~\cite{yuan2019simple}. 
%Further increasing CNN layers or $k$ does not yield significantly better results for these datasets.
%\footnote{\scriptsize Note that the hidden dimension $k$ and the number of dilated convolutional layers highly depend on the datasets. In our later work we find that the optimal performance is obtained when $k$ is up to 1024 and the number of convolutional layers is up to 40 in the Movielens~\cite{yuan2020future,tang2018caser} datasets with sequence size of 100. } 
Batch size $b$ and kernel size are set to 512  and 3  respectively for all models. 

As for the pre-trained model, we use  90\% of the dataset in $\mathcal{S}$ for training, and the remaining for validation. Different from fine-tuning, the measures for pre-training (i.e., MRR@5) are calculated based on the rank in the whole item pool following~\cite{yuan2019simple}.
We use  $\eta$ =  0.001 for all pre-trained models. Batch size is set to 32 and 128 for causal and non-causal CNNs due to the consideration of memory limitation.  The masked percentage for non-causal CNNs is 30\%. Other parameters are kept the same as mentioned above.

\subsection{RQ1.}
\label{RQ1}
Since PeterRec has a variety of variants with different circumstances (e.g., causal and non-causal versions, different insert methods (see Figure~\ref{modelpatch}),  and different fine-tuning architectures (see Figure~\ref{finetunearch})), presenting all results on all the  five datasets  is redundant and space unacceptable.   Hence, in what follows, we report parts of the results with respect to some variants of PeterRec (on some datasets or metrics)  considering that their behaviors are consistent.

To answer RQ1, we report the results in Figure~\ref{impactofpre-training} \& Table ~\ref{finezeroandfineall}.  For all compared models, we use the causal CNN architecture. For PeterRec,
we use the serial insertion in Figure~\ref{modelpatch} (c). First, we observe that PeterRec  outperforms PeterZero with large improvements on all the five datasets. Since PeterRec and PeterZero use exactly the same  network architectures and  hyper-parameters,  we can draw the conclusion that the self-supervised pre-trained user representation is of great importance in  improving the accuracy of  downstream tasks. To further verify it, we also report results of FineAll and FineZero in  Table ~\ref{finezeroandfineall}. Similarly,
FineAll largely exceeds FineZero (i.e., FineAll with random initialization) on all datasets. The same conclusion also applies to FineCLS  and FineLast with their random initialization variants.
%which  re-trains all parameters of the fine-tuning model by random initialization ---  e.g., a normal distribution between 0$\sim $1.

%Note only PeterRec, other models FIneAll, FineCLS, FineLast gives exactly the same observations.

\begin{figure}
	\small
	%	\vspace{-0.1in}
	\centering     %%% not \center
	\subfigure[\scriptsize Cold-Rec2  (one epoch:  $500*b$) ]{\label{yahoo-alphazero}\includegraphics[width=0.235\textwidth]{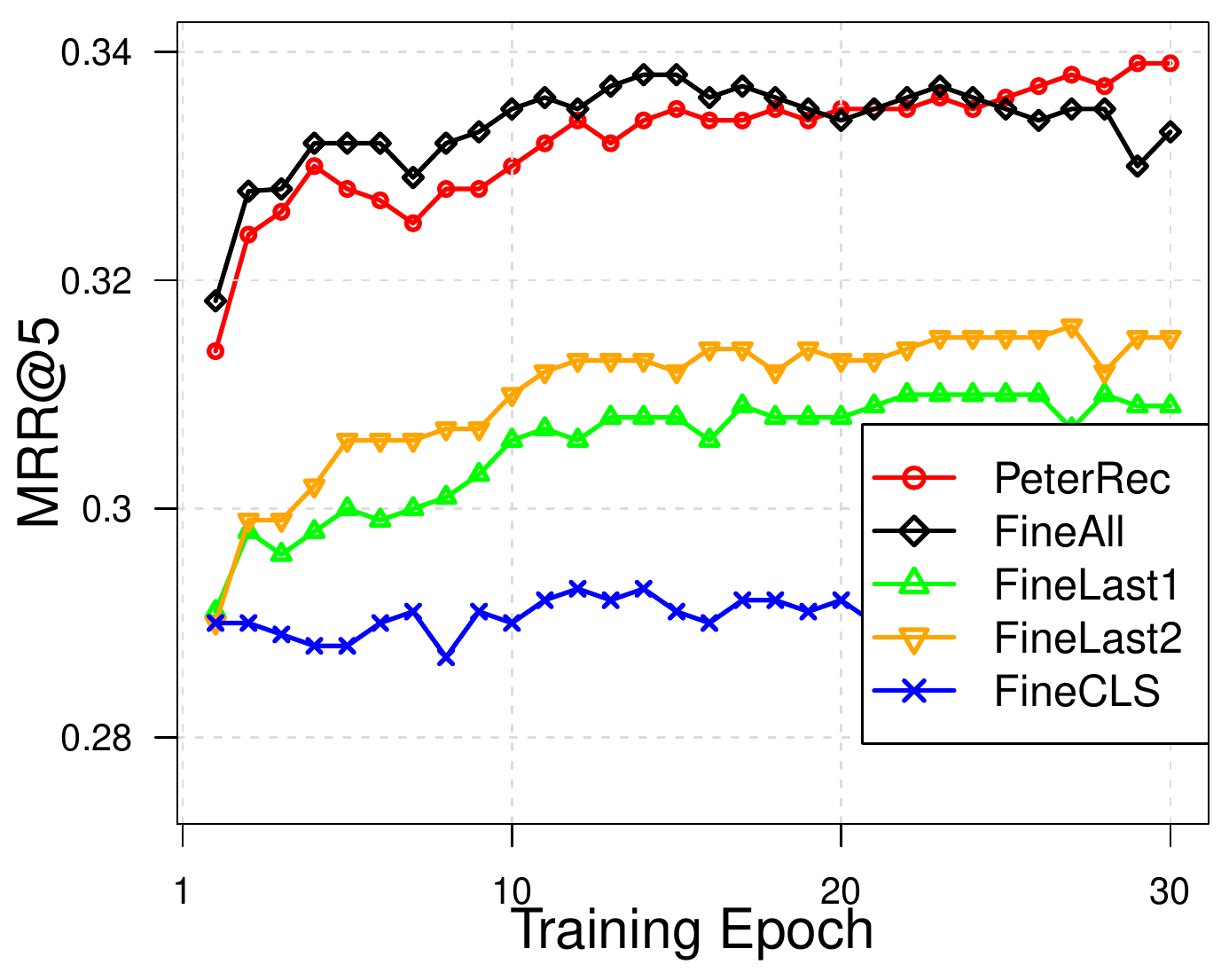}}
	\subfigure[\scriptsize AgeEst ( one epoch:  $500*b$ ) ]{\label{yahoo-alpha}\includegraphics[width=0.235\textwidth]{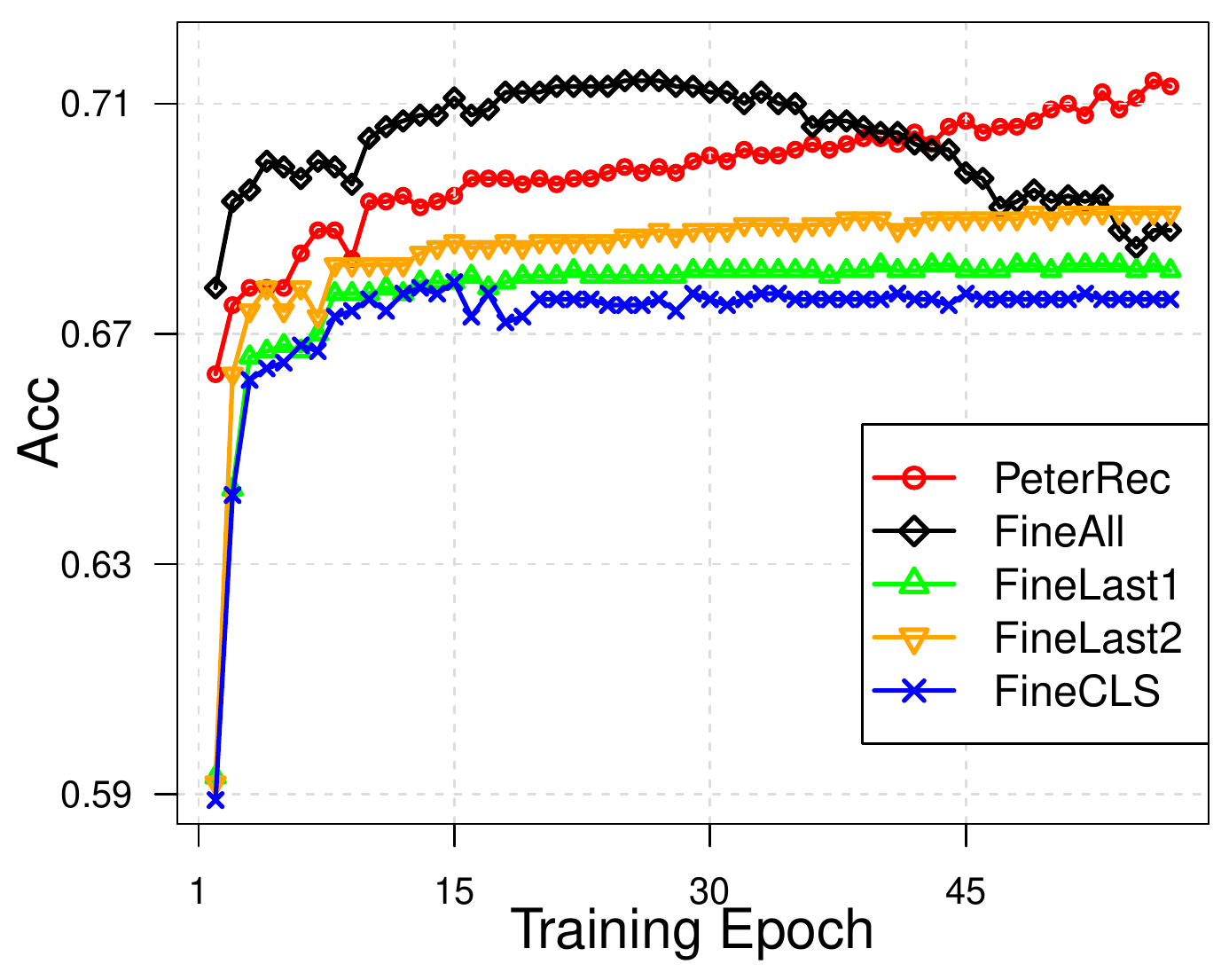}}
	%	\subfigure[\scriptsize  HR@5]{\label{yahoo-alpha}\includegraphics[width=0.24\textwidth]{./image/bertimage/yahoo-finetune.pdf}}
	%	\subfigure[\scriptsize  NDCG@5]{\label{yahoo-alpha}\includegraphics[width=0.24\textwidth]{./image/bertimage/yahoo-finetune.pdf}}
	\caption{\small Convergence behaviors of PeterRec and baselines  (with the non-causal CNN). FineLast1 and  FineLast2 denote FineLasts that optimize  only the last one and two CNN layers (including the corresponding layer normalizations), respectively. All models here have fully converged. The number of parameters to be re-learned: FineAll$\gg$ FineLast2$>$ PeterRec$ \approx $FineLast1$>$FineCLS.} 
	\label{convergencebeh}
\end{figure}

\subsection{RQ2.}
\begin{figure}
	\small
	%		\vspace{-0.1in}
	\centering     %%% not \center
	\subfigure[\scriptsize 5 \% of training data (one epoch:  $500*b$) ]{\label{yahoo-alphazero}\includegraphics[width=0.235\textwidth]{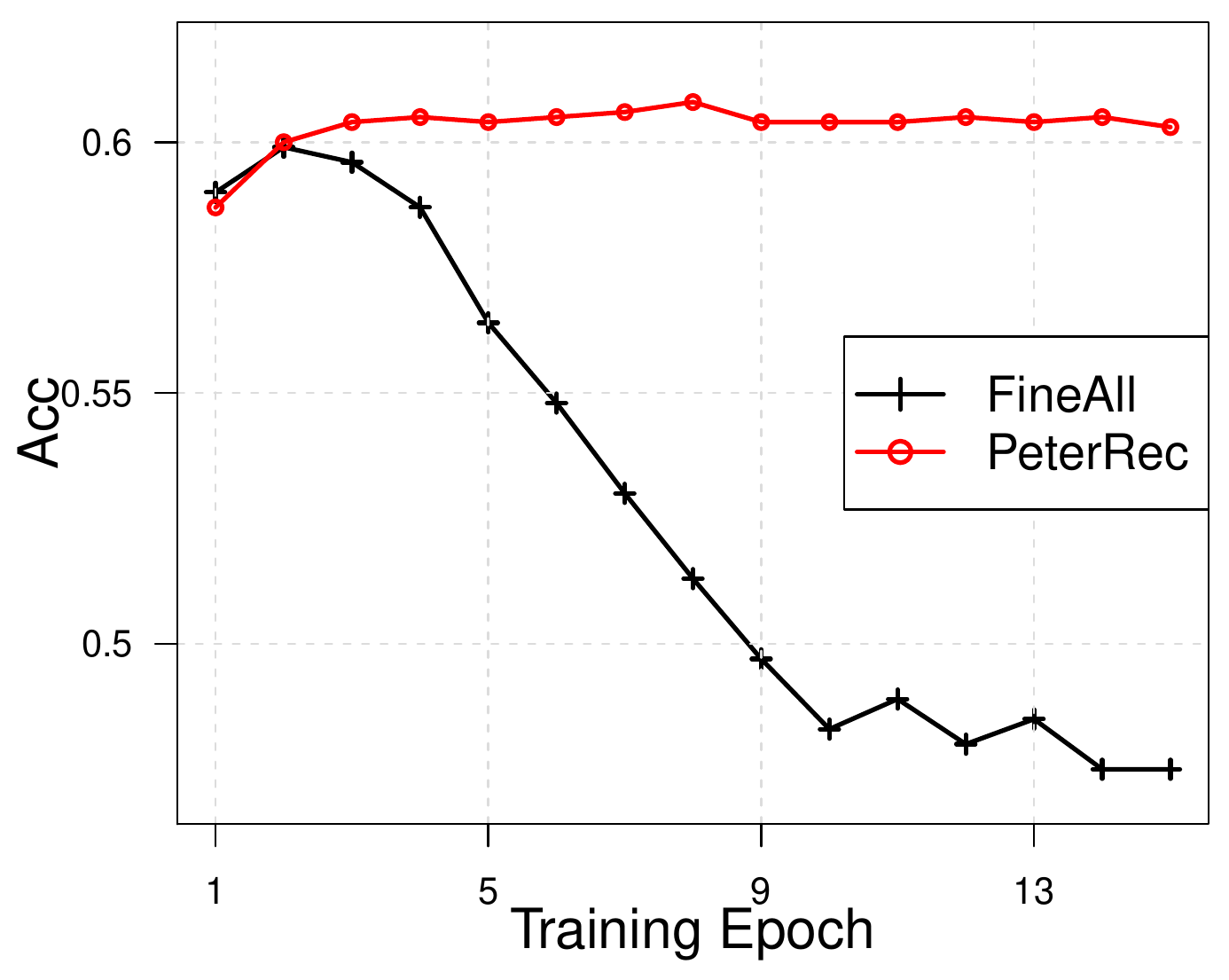}}
	\subfigure[\scriptsize 10 \% of training data (one epoch:  $500*b$) ]{\label{yahoo-alpha}\includegraphics[width=0.235\textwidth]{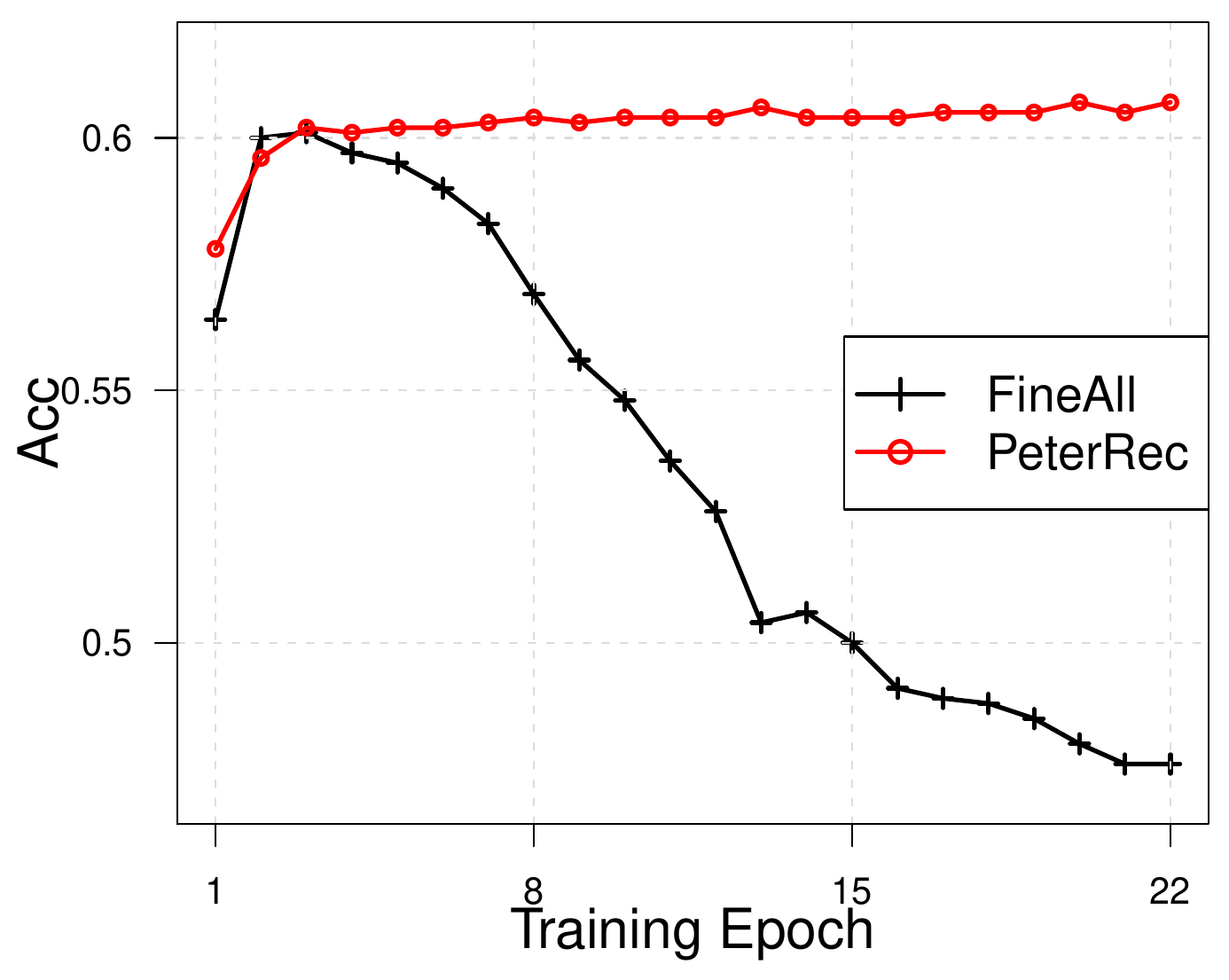}}
	\caption{\small Convergence behaviors of PeterRec and FineAll on LifeEst using much less training data. The improvements of PeterRec relative to FineAll is around 1.5\% and 1.7\%  on  (a) and (b) respectively  in terms of the optimal performance.} 
	\label{convergewithlimiteddata}
\end{figure}

To answer RQ2, we  report the results in Table~\ref{performcmp}.  We use  the non-causal CNN architecture for all models and parallel insertion for PeterRec.
First, we observe that with the same pre-training model, FineCLS and FineLast perform much worse than FineAll, which demonstrates that fine-tuning the entire model benefits more than tuning only the last (few) layers. Second, we observe that
PeterRec achieves  similar results with FineAll, which suggests that  fine-tuning the proposed model patch (MP) is as effective as fine-tuning the entire model. By contrast, PeterRec retains most pre-trained parameters (i.e.,  $\tilde{\Theta}$) unchanged for any downstream task, whereas FineAll requires a large separate set of parameters to be re-trained and saved for each task, and thus is not efficient for resource-limited applications and multi-domain learning settings. Moreover, fine-tuning all parameters may easily cause the overfitting (see Figure~\ref{convergencebeh} (b) and Figure~\ref{convergewithlimiteddata}) problems. To clearly see the convergence behaviors of these models, we also plot their results on the ColdRec and AgeEst datasets in Figure~\ref{convergencebeh}.
\begin{table} 
	\centering
	%	\vspace{-0.1in}
	\caption{\small Results regarding user profile prediction.  
		%		 Interestingly, we find that the number of male users is much larger than females with the ratio of 72.5\% : 27.5\% in the ground truth of GenEst. 
	}
	\small
	%	\vspace{-0.1in}
	\label{CLScom}
	\setlength{\tabcolsep}{5mm}
	\begin{threeparttable}				
		\begin{tabular}{c|c|c|c}
			\toprule
			\small Model &   \small GenEst&  \small AgeEst&  \small LifeEst\\
			\midrule
			%			FineZero        &0.304 &0.290  &0.893& 0.703& 0.596\\ 	
			LabelCS      &0.725  &0.392 &  0.446\\ 	
			\midrule
			MTL  & 0.899& 0.706& 0.599\\ 
			\midrule
			PeterRec  & 0.906& 0.714& 0.615\\ 							
			\bottomrule
		\end{tabular}
	\end{threeparttable}
	%		\vspace{-0.1in}
\end{table}

\begin{table} 
	\centering
	\caption{\small Top-5 Accuracy in the cold user scenario. 
		%		We report ColdRec-1 and  ColdRec-2 with HR@5 and MRR@5, respectively. 
		%		NeuFM and DeepFM use one and three fully connected layer, respectively,  and $k$ is set to 256 to obtain the optimal performance.
	}
	\small
	%		\vspace{-0.05in}
	\label{coldusercomp}
	\setlength{\tabcolsep}{3.9mm}
	\begin{threeparttable}				
		\begin{tabular}{c|c|c|c|c}
			\toprule
			\small Data &   \small NeuFM&  \small DeepFM& \small MTL &  \small PeterRec\\
			\midrule
			%			FineZero        &0.304 &0.290  &0.893& 0.703& 0.596\\ 	
			ColdRec-1      &0.335  &0.326& 0.337&0.351\\ 	
			\midrule
			ColdRec-2  & 0.321& 0.317 &0.319  &0.339\\ 							
			\bottomrule
		\end{tabular}
	\end{threeparttable}
	%		\vspace{-0.1in}
\end{table}

\subsection{RQ3.}
\vspace{0.05in}
To answer RQ3, we demonstrate the results in Table~\ref{CLScom} and~\ref{coldusercomp}. Clearly, PeterRec notably outperforms LabelCS, which demonstrates its the effectiveness in estimating user profiles. Meanwhile, PeterRec  yields better top-$5$ accuracy than NeuFM, DeepFM and MTL  in the cold-user item recommendation task. Particularly, PeterRec outperforms MTL in all tasks, which implies that the proposed two-stage pre-training \& fine-tuning paradigm  is more powerful than the  joint training  in MTL. We argue this is because  the optimal parameters  learned for two objectives in MTL does not gurantee  optimal performance for
fine-tuning.
%Interestingly, MTL performs much worse than NeuFM and DeepFM despite with more training objectives. 
%We argue this is because the optimal parameters  learned for the two tasks may be non-optimal for one of them. 
%We also observe that MTL does not have the overfitting problem compared with NeuFM and DeepFM since the  training loss for the source task works as an additional regular for that of the target task. 
Meanwhile, PeterRec is able to take advantage of all
training examples in the upstream task, while these baseline models only leverage traing examples that
have the same users involved in the target task. Compared with these baselines,
%mention overfitting in arxiv
%Moreover, PeterRec is a general and flexible model which can be directly applied for various downstream tasks, whereas NeuFM and DeepFM are specifically designed for the item recommendation task. In addition, similar to the above analysis, 
PeterRec is memory-efficient since it only maintains a small set of model patch parameters
for a new task while  others have to store all parameters for each task.
% with much more parameters. 
In addition, the training speed of MTL is several times slower than PeterRec due to the expensive pre-training objective functions.
If there are a large number of sub-tasks, PeterRec will always be a better choice considering its high degree of parameter sharing. 
To the best of our knowledge, PeterRec is the first model that considers the memory-efficiency issue for multi-domain recommendations. 

\begin{table} 
	\centering
	\caption{\small PeterRecal vs. PeterRecon. The results of the first and last two columns are  ColdRec-1 and AgeEst datasets, respectively.  }
	\small
	%		\vspace{-0.05in}
	\label{prefinecompare}
	\setlength{\tabcolsep}{2.05mm}
	\begin{threeparttable}				
		\begin{tabular}{c|c|c||c|c}
			\toprule
			\small Model &  \small PeterRecal& \small PeterRecon &  \small PeterRecal&  \small PeterRecon\\
			\midrule
			%			FineZero        &0.304 &0.290  &0.893& 0.703& 0.596\\ 	
			Pretaining       &0.023   &   0.020 & 0.045 & 0.043\\ 	
			\midrule
			Fine-tuning        &0.348 &0.351  &0.708&0.714 \\ 							
			\bottomrule
		\end{tabular}
	\end{threeparttable}
	%	\vspace{-0.1in}
\end{table}

\begin{table} 
	\centering
	\caption{\small Performance of different insertions in Figure~\ref{modelpatch} on AgeEst.   }
	\small
	%	\vspace{-0.05in}
	\label{insertionmethod}
	\setlength{\tabcolsep}{4.8mm}
	\begin{threeparttable}				
		\begin{tabular}{c|c|c|c|c}
			\toprule
			\small Model &  \small (b) & \small (c) &  \small (d) &  \small \ (e)\\
			\midrule
			PeterRecal        & 0.708 &0.708  &0.708&0.675 \\ 	 		
			\midrule
			%			FineZero        &0.304 &0.290  &0.893& 0.703& 0.596\\ 	
			PeterRecon        &0.710   &0.710  & 0.714&0.685\\ 	
			\bottomrule
		\end{tabular}
	\end{threeparttable}
	%	\vspace{-0.1in}
\end{table}

\subsection{RQ4.}
\label{rq4}
This subsection offers several insightful findings: (1) By contrasting PeterRecal and PeterRecon in Table~\ref{prefinecompare}, we can draw the conclusion that better pre-training models for sequential recommendation may not necessarily lead to better transfer-learning accuracy. This is probably because PeterRecon takes two-side contexts into consideration~\cite{yuan2020future}, which is more effective than the sequential patterns learned by PeterRecal for these downstream tasks. However, for the same model, better pre-training models usually lead to better fine-tuning performance. Such results are simply omitted due to limited space. (2) By comparing results in Table~\ref{insertionmethod}, we observe that for parallel insertion, the MP  has to be inserted before the normalization layer. We argue that the parallelly inserted MP in Figure~\ref{modelpatch} (e) may break up the addition operation in the original residual block architecture (see FIgure~\ref{modelpatch} (a)) since MP in (e) introduces two additional summation operations, including the sum in MP and sum with the ReLU layer.  (3) In practice, it is usually very expensive to collect a large amount of user profile data, hence we present the results with limited training examples in Figure~\ref{convergewithlimiteddata}. As clearly shown, with limited training data, PeterRec performs better than FineAll, and more importantly, PeterRec is very stable during fine-tuning since only a fraction of  parameters are learned. By contrast, FineAll has a severe overfitting issue, which cannot be solved by regularization or dropout techniques.

\section{Conclusions}
In this paper, we have shown that (1) it is possible to learn universal user representations by modeling only unsupervised user sequential behaviors; and (2)  it is also possible to adapt the learned representations for a variety of downstream tasks. By introducing the grafting model patch, PeterRec allows all pre-trained parameters unchanged  during fine-tuning, enabling efficent \& effective adaption to multiple domains with
only a small set of re-learned  parameters for a new task.  We have evaluated several alternative designs of PeterRec, and made insightful observations by extensive ablation studies. 
By releasing both high-quality datasets and codes, we hope PeterRec serves as a benchmark for transfer learning in the recommender system domain.

We believe PeteRec can be applied in more domains  aside from tasks in this paper. For example, if we have
the video watch behaviors of a  teenager, we may know whether he has depression or propensity for violence  by PeterRec without resorting to much feature engineering and  human-labeled data. This can remind  parents  taking measures in advance to keep their children free from such issues.
For future work, we may explore PeteRec with more tasks.
\section*{Acknowledgement}
This work is partly supported by the National Natural Science Foundation of China (61972372, U19A2079).
%We would like to Beibei Kong, Jian Xiong and the Oula team at Tencent for the help in generating the original datasets.

%%
%% The next two lines define the bibliography style to be used, and
%% the bibliography file.
\bibliographystyle{ACM-Reference-Format}
\bibliography{bibliography}

%%
%% If your work has an appendix, this is the place to put it.
\appendix

\end{document}